# Combined Daily Monitoring of Aerosol Optical Depths and Water Vapour Column Content during LACE 98 and LITFASS 98 Experiments.

VICTOR NOVIKOV[2)], ULRICH LEITERER[1)], GALINA ALEKSEEVA[2)], VIATCHESLAV GALKIN[2)], JÜRGEN GÜLDNER[1)], AND TATJANA NAEBERT.[1)],

1) German Weather Service, Meteorological Observatory, 15848 Tauche OT Lindenberg, Germany

2) Russian Academy of Sciences, Pulkovo Observatory, 196140 St. Petersburg, Russia

E-mail: novikov_victor@.mail.ru , novikov@gao.spb.ru ;

Ulrich.Leiterer@dwd.de ;

alekseeva@gao.spb.ru ;

galkin@gao.spb.ru ;

Juergen.Gueldner@dwd.de ;

Tatjana.Naebert@dwd.de

**Summary**

During summer of 1998 two large-scale complex campaigns, LITFASS98 (May 25th to June 22nd) and LACE98 (July 13th to August 14th), took place at the Meteorological Observatory Lindenberg (MOL). The aim of both experiments focus on the intensive daily observations of atmospheric conditions and the determination of their fundamental meteorological parameters in the vertical column over Lindenberg (Lindenberg's Column). About 20 German research institutions and addition one from the Netherlands, Austria and Russia participated at the experiments. A wide variety of ground-based instruments was operated in Lindenberg and Falkenberg, including LIDARs, microwave radiometer and radiosondes complemented by tethered balloons and aircraft measurements. For the first time the star- and sunphotometer of MOL were used together with other geophysical tools. The observations with both photometers were carried out practically every day and night except during absolutely overcast conditions. The observed data were processed immediately by a series of programs developed at Pulkovo Observatory (Russia), and the results (daily variations of aerosol optical depths and water vapour column content) were presented at daily briefings. The comparison of these results with radiosonde and microwave radiometer data demonstrated the usefulness of photometer data for the calibration of other ground-based observations and satellite measurements.

## 1 Used Instruments and Method

We used the automatic sunphotometer (ROBAS 30) for sun-observations. For star-observations we used the new semi-automatic version of a starphotometer built in co-operation by MOL, Dr. Schulz & Partner GmbH and Pulkovo Observatory in 1997. Both photometers have the same set of quasi-monochromatic interference filters. For the determination of **aerosol optical depths** we used the following **wavelengths (in µm)**:

**0.3991, 0.4512, 0.4922, 0.5510, 0.6538, 0.7915, 0.8553, 1.0539 (sunphotometer);**

**0.3907, 0.4444, 0.5007, 0.5328, 0.6019, 0.6729, 0.7786, 0.8629, 1.0452 (starphotometer).**

For the determination of the **water vapour column** content the filters **0.9517** and **0.9446** were used (for sun- and starphotometry respectively).

The transmission curves for all filters were determined in the optic laboratory of MOL. The time-interval between the observations is 10 min for sunphotometry and about 3-5 min for starphotometry. The methods used are described in detail by LEITERER et al., [1998]. One can find the main list of references in that paper too.

Here we present only main equations or expressions with a short method description.

$$m_\lambda = -2,5 \log U_\lambda \qquad (1)$$

with

$U_\lambda$ - signal from object measured (voltages or impulses/sec) in wavelength $\lambda$

$$m_\lambda^{obs.} = m_\lambda^{o} + \alpha_\lambda \cdot F(z) \qquad \text{(Bouger-Lambert' law)} \qquad (2)$$

with $m_\lambda^{obs.}$ - observed star magnitude

$m_\lambda^{o}$ - extraterrestrial instrumental star magnitude

$\alpha_\lambda$ - total extinction for air mass **one**

$F(z)$ - air mass

The astronomical extinction coefficient $\alpha_\lambda$ and the meteorological total optical thickness $\delta_\lambda$ are connected by the expression:

$$\alpha_\lambda = (-2,5 / \ln 10) \cdot \delta_\lambda = 1.085736\, \delta_\lambda \qquad (3)$$

The coefficient of extinction $\alpha_\lambda$ in the Equation **(2)** includes continuous components of extinction and equals:

$$\alpha_\lambda = \alpha_{\lambda,Ray} + \alpha_{\lambda,aer} + \alpha_{\lambda,ozone} + \alpha_{\lambda,NO2} \qquad (4)$$

with

$\alpha_{\lambda,Ray}$, $\alpha_{\lambda,aer}$, $\alpha_{\lambda,ozone}$ ,$\alpha_{\lambda,NO2}$- Rayleigh, aerosol, ozone, and nitrogen dioxide components of extinction accordingly. If we have calculated $\alpha_{\lambda,Ray}$ and obtained $\alpha_{\lambda,ozone}$ ,$\alpha_{\lambda,NO2}$ from independent sources, it's possible to determine $\alpha_{\lambda,aer}$ resp. $\delta_{\lambda,aer} = \alpha_{\lambda,aer} / 1.085736$:

$$\alpha_{\lambda,aer} = \alpha_\lambda - (\alpha_{\lambda,Ray} + \alpha_{\lambda,ozone} + \alpha_{\lambda,NO2}) \qquad (5)$$

For the water vapour band the Equation **(2)** is extended to

$$m_\lambda^{obs.} = m_\lambda^o + \alpha_\lambda \cdot F(z) + \Delta m\,(W) \qquad (6)$$

with

$\Delta m\,(W)$ - water vapour absorption (in star magnitudes)

**W** - water vapour content on the line of view (in **cmppw**)

For the absorption magnitude (in the water vapour band for the filter wavelengths 0,9517 and 0,9446 μm) we used an empirical model developed by GOLUBITSKY and MOSKALENKO, 1968 and GALKIN and ARKHAROV, 1981.

$$\Delta m\,(W) = C_\lambda \cdot [W_o \cdot F(z)]^{\mu_\lambda} \qquad (7)$$

where $C_\lambda$ and $\mu_\lambda$ are empirical parameters, $W_o$ – denotes the water vapour content in the vertical atmospheric column [in cmppw]. The spectral parameters $C_\lambda$ and $\mu_\lambda$ received from laboratory investigations in Pulkovo Observatory (ALEKSEEVA et al., 1994) may be used for the calculation of the integral parameters for any filter with a known spectral transmission curve.

In order to use Equation **(2)** for the determination of $\alpha_\lambda$ directly we must know the individual spectral instrumental extraterrestrial magnitudes $m_\lambda^o$ for the selected star.

If there is only one source of radiation (sun), we have only one way to receive such magnitudes, namely to execute the set of observations in a region with very stable atmospheric extinction ($\alpha_\lambda$ = const.). For the sunphotometer such observations were carried out at Zugspitze (2970 m above sea level, German Alps), and individual spectral solar extraterrestrial magnitudes were determined for all filters. One must take into account that the solar magnitude of the „water“ filter will be falsified (the application of the Bouger-Lambert' law is incorrect in this case).

If we have many sources of radiation (star observations) we may use the so-called **Δ–method** (or Two Star Differential Method) as **the first approximation** for $\alpha_\lambda$-determinations. For two stars with ΔF(z)≥1 we get from Equation **(2)**:

$$\alpha_\lambda = \frac{\left({}^{2}m_\lambda^{obs.} - {}^{1}m_\lambda^{obs.}\right) - \left({}^{2}m_\lambda^{0} - {}^{1}m_\lambda^{0}\right)}{F_2(z) - F_1(z)} \quad \textbf{(8)}$$

In this case it is not necessary to know the individual extraterrestrial instrumental magnitudes. It is sufficient to know only the difference between the extraterrestrial magnitudes for these stars. This difference may be obtained from any homogeneous spectrophotometric catalogue because for monochromatic (or quasi-monochromatic) radiation this difference is constant for whichever system. We used the Pulkovo Spectrophotometric Catalogue (ALEKSEEVA et al., 1996) as it is most homogeneous.

When the first values $\alpha_\lambda$ (for mean moments between two stars observations) are acquired according to Equation **(8)**, we can determine the time-dependencies of $\alpha_\lambda$ (UTC) for all filters during one night by means of polynomial approximation. Then the individual $\alpha_\lambda$-values may be calculated from these time-dependencies, and the individual instrumental extraterrestrial magnitudes $\mathbf{m_\lambda^o}$ may be computed for every star observation. And finally **mean** individual star magnitudes $\overline{m_\lambda^0}$ may be calculated for every individual star used. These $\overline{m_\lambda^0}$ may be used directly for $\alpha_\lambda$-determinations from Equation **(2)**. The procedure described above we call **“the second approximation“**.

For **water vapour content** determinations the procedure is more complicated. In order to use the Equations **(6)** and **(7)** for $\mathbf{W_0}$-determinations we have to know not only $\mathbf{m_\lambda^o}$, but the parameters $\mathbf{C_\lambda}$ and $\boldsymbol{\mu_\lambda}$, too. As the first approximation we use the parameters calculated from spectral laboratory data, ALEKSEEVA et al., 1994, both for sun- and starphotometer „water“ filters.

For **sun observations** we are forced to use the “falsified” value $\mathbf{m_\lambda^o}$ as **the first approximation** received according to the Bouger-Lambert' law. Then we get the first preliminary values $W_0$ using the Equations **(6)** and **(7)**.

For **star observations** we can get the first values $\mathbf{W_0}$ using the **Δ-method**:

$$\mathbf{W_0} = \left[ \frac{\left({}^{2}\mathrm{m}_{\lambda}^{\mathrm{obs.}} - {}^{1}\mathrm{m}_{\lambda}^{\mathrm{obs.}}\right) - \left({}^{2}\mathrm{m}_{\lambda}^{0} - {}^{1}\mathrm{m}_{\lambda}^{0}\right) - \alpha_{\lambda} \cdot \left(\mathrm{F}_2(z) - \mathrm{F}_1(z)\right)}{\mathbf{C}_{\lambda} \cdot \left(\mathrm{F}_2(z)^{\mu} - \mathrm{F}_1(z)^{\mu}\right)} \right]^{\frac{1}{\mu}} \qquad \mathbf{(9)}$$

$\alpha_\lambda$ for „water“ filters are calculated by logarithmic interpolation between the nearest filters outside the water vapour band or by power approximation of data from all other filters.

Then we can carry out „**the second approximation**“ for $\mathbf{W_0}$ and receive **mean** star magnitudes $\overline{\mathrm{m}_{\lambda}^{0}}$ in the „water“ filter channel for all stars using Equations **(6)** and **(7)**.

With the obtained **mean** individual star magnitudes $\overline{\mathrm{m}_{\lambda}^{0}}$ and with „laboratory“ parameters $\mathbf{C_\lambda}$ and $\boldsymbol{\mu_\lambda}$ all data of the star observations were recalculated. Then we executed the calibration procedure using radiosondes $\mathbf{W_0}$ -data in order to define „real“ parameters $\mathbf{C_\lambda}$ and $\boldsymbol{\mu_\lambda}$, which agree best. The calibration curve for all observations by **starphotometer** is shown in **Figure 1**. It is an exponential approximation of the dependence $\mathbf{\Delta m\ (W)}$ where $\mathbf{W = W_{0RS80}}$ P $\mathbf{F(z)}$ is the water vapour content on the line of view (in **cmppw**). We obtained the parameters

$\mathbf{C_\lambda = 0.^{m}582}$ and $\boldsymbol{\mu_\lambda} \mathbf{= 0.548}$ **starphotometer (10)**

For the **sunphotometer** we used the other dependence:

$$\mathbf{m_\lambda^{obs.} - (\alpha_{\lambda,Ray} + \alpha_{\lambda,Aer}) \cdot F(z) = C_\lambda \cdot [W_{0RS80} \cdot F(z)]^{\mu_\lambda} + m^{o}_\lambda} \qquad \mathbf{(11)}$$

The left side of Equation **(11)** is known from observations. The parameter $\boldsymbol{\mu_\lambda}$ at the right side of Equation **(11)** was determined from Pulkovo „laboratory“ data (ALEKSEEVA et al., 1994). Hence, by constructing the most probable line, we can receive the extraterrestrial instrumental solar magnitude $\mathbf{m_\lambda^{o}}$ at the point of intersection with the ordinate. The inclination of the straight line gives the value of parameter $\mathbf{C_\lambda}$ specified on RS80-data, which are specially calibrated and corrected as described by Leiterer et al., 2000.

This calibration curve for all observations by **sunphotometer** is shown in **Figure 2**. Following values are the result:

$$C_\lambda = 0.^m4714\,, \quad \mu_\lambda = 0.5805 \text{ and } m_\lambda^o = -4.7228 \qquad \textbf{sunphotometer} \qquad \textbf{(12)}$$

All observed data for both photometers were recalculated with $C_\lambda$ and $\mu_\lambda$ parameters of the expressions **(10)** and **(12)**, and the final values of $W_0$ [cmppw] were obtained.

Microwave radiometer WVR-1100 (F = 23.8 GHz on, 31.4 GHz off; sample time 1 s, cycle time < 1 min; running mean values of 10 min) was also used for monitoring of column precipitable water **PW** ($\equiv W_0$). More details are described by GÜLDNER AND SPÄNKUCH, 1999.

## 2 Results and Discussion

We present the $\delta_{\lambda,aer}$- and **PW**-monitoring data for the periods of May 26$^{th}$ to June 22nd and July 14$^{th}$ to August 16th.

The set of $\delta_{\lambda,aer}$-data consists of individual spectral values for all „aerosol channels“ and interpolated data for the wavelengths $\lambda = 0.550$ and 1.00 μm obtained by exponential approximation

$$\delta_{\lambda,aer} = \beta \cdot \lambda^{\text{Å}} \qquad \textbf{(13)}$$

where wavelength $\lambda$ is given in μm, $\beta$ is the aerosol optical depth at 1.0 μm, Å is the “Angstrom” exponent. Usually, the Angstrom exponent is denoted with $\alpha$, but in this paper the letter $\alpha$ has to be used for the astronomical extinction coefficient, see Equation **(3)**.

For „water channels“ the data can be obtained by three different methods:

- radiosonde (RS80, A-Humicap) relative humidity profiles, every 6 hours; specially calibrated as reported by Leiterer et al., 2000.
- optical methods (according to Equations **1-12**) for sun- and starphotometry;
- microwave radiometer data (running mean values of 10 min).

The mean accuracy of sun- and starphotometry both for $\delta_{\lambda,aer}$- and **PW**-data is about 2-3% (for conditions with very unstable extinction it may decrease to 5-7%). The detailed analysis of accuracy is presented in the final report, Leiterer et al., October 2000, of the research project 436 RUS 113/76 funded by the Deutsche Forschungsgemeinschaft.

All monitoring data are accessible as Excel 5.0-files in MOL (U. Leiterer, E-mail: ulrich.leiterer@dwd.de).

The time series of the observed data are shown in **Figures 3, 4** and **Figures 5, 6** (for **PW**- and $\delta_{\lambda,\mathbf{aer}}$-data respectively). The details of the comparison for some selected dates are shown in **Figures 7** to **11**. Some remarkable features of aerosol behaviour are shown in **Figures 12** to **15**. For a more correct comparison with sunphotometric and microwave radiometer data we used 3-point-smoothing for starphotometric data. For some periods there are no sun- or starphotometer observations caused by thick clouds. Also, for some dates there are no microwave radiometer measurements due to technical problems.

One can see that as a rule the **PW**-data obtained by the three methods agree very well. The mean differences are no bigger than 5%, only sometimes the differences increase to 10%.

The important feature is that the individual **PW**- time-variations have the same form for microwave radiometer, sun- and starphotometer data. That means we can systematically investigate the real rapid **PW**- time-variations.

Now we describe both $\delta_{\lambda,\mathbf{aer}}$- and **PW**-behaviour during all monitoring periods, in more detail.

## 2.1 Details in water vapour column (PW) and aerosol optical depth ($\delta_{Ae}$) monitoring

Looking at the observation periods of the LITFASS (May/June 1998, **Figures 3** and **5**) and LACE (July/August 1998, **Figures 4** and **6**) experiments you get the following picture.
The water vapour column **PW** changes according to the weather conditions in a range of 1.0 and 3.7 cmppw.
The aerosol optical depth $\delta_{\mathbf{Ae}}$ ranges from 0.01 to more than 1.0 for the visible as well as for the infrared spectrum at $\lambda$ = 0.5 respectively 1.0 µm.
The variability of the two variable constituents (water vapour and aerosol) is therefore considerable, that is, approximately factor 4 for the water vapour column and factor 50 for the aerosol optical depth..

The water vapour record is without gaps due to the combined use of the 3 measuring methods (radiosonde, microwave radiometer and sun-/starphotometry). The radiosonde observations in an interval of 6 hours play an important role here.

The aerosol content of the atmospheric column above Lindenberg can only be measured at undisturbed sun- or starlight using optical methods. Therefore measuring is often interrupted by clouds. Low **PW**-values are very often connected with low $\delta_{Ae}$-values. As an example (see **Figures 9** and **10**) we can take the decrease in the **PW**-values of ca. 3.5 cmppw on August 7$^{th}$ to approximately 2.5 cmppw on August 8$^{th}$ and to only 1.2 cmppw on August 10$^{th}$. The $\delta_{Ae}$-values for a wavelength $\lambda$ = 0.55 µm decreased as follows: from approx. 0.25 on August 7$^{th}$ to approx. 0.15 on August 8$^{th}$ evening and from 0.12 on August 10$^{th}$ during the day to a minimum of 0.03 in the following night.

The curve of the **PW**-value does not necessarily correlate with the curve of the $\delta_{Ae}$-value (see **Figures 12** and **13**). In the night of June 21$^{st}$/22$^{nd}$ the **PW**-values remain almost unchanged with approx. 2.3 cmppw, where as the $\delta_{Ae}$-values for a wavelength $\lambda$ = 0.5 µm tripled from approx. 0.3 to 0.9 (detail in **Figure 12**, time period 22-01 UTC) caused by the formation of humid haze.

Likewise a decrease of $\delta_{Ae}$ for the wavelength $\lambda$ = 0.5 µm from 0.12 to only 0.01 is possibly within one hour (23-00 UTC, **Figure 13**) combined with a relative modest decrease of the **PW**-values of only 0.3 cmppw from 1.6 to 1.3 cmppw (microwave-radiometer data) or from 2.0 to 1.7 cmppw (starphotometer data). In this case one has also some negative bias of microwave-radiometer to starphotometer or radiosonde data as discussed in the next section.

Although relative humidity does influence the growth of aerosol particles, especially at relative humidities bigger than 60 %, the absolute humidities (and with that the **PW**-values ) and the optical aerosol depths $\delta_{Ae}$ are mutually independent variable atmospheric constituents.

The absolute humidities mainly depend on temperature and are characterised by the geographical origin of the air mass (land/sea). The optical aerosol depths that cause the turbidity of the atmosphere are not dependent on temperature but are also influenced by their geographical origin (ways of transport and sources of emission).

## 2.2 Details in water vapour column comparison

If you compare the three methods for the estimation of the water vapour column each of the 3 methods

- radiosonde
- microwave radiometry
- sun- and starphotometry (optical method)

reflect the trend and the value of the **PW**-value very well.

The microwave radiometer has been calibrated with a climatological data record (years 1991-94) of the Lindenberg radiosonde ascents, merely for some meteorological situations there are deviations. Thus **Figure 7** (June 4th, 6:00 UTC to June 5th, 11:00 UTC) shows a very good agreement of all three methods within approx. ±0,1 cmppw. But in the evening of June 5th between 17:00 and 23:00 UTC considerable differences between all three methods appear. At 17:00 UTC the radiosonde value at 2.0 cmppw is clearly below the sunphotometer value at 2.4 cmppw and the microwave value agree at 2.9 cmppw, but the starphotometer value is clearly lower at 2.7 cmppw. The radiosonde values still are the statistical basis of calibration for the microwave method and the optical methods. However, caused by the drift of the radiosonde the radiosonde value may in individual cases differ considerably from the measured microwave radiometer value or the value obtained with optical methods which refers to the air mass above the observation point. Lower PW-values were detected more often with the microwave radiometer method than with the radiosonde, respectively “optical” methods, see the details in **Figures 8, 9, 13**. The physical background for the occasional differences, especially between “optical” and microwave radiometer methods, has not been sufficiently investigated and should be the subject of future validating experiments. However, the advantages of applying physically completely different measuring principles in order to determine the water vapour column become clear. Only in this way it is possible to recognise the advantages and the weak points of the different measuring methods in order to achieve tangible improvements.

## 2.3 Details in monitoring of the spectral optical aerosol depth $\delta_{Ae}$, LIDAR comparison

The details of the spectral characteristic of the optical aerosol depth $\delta_{Ae}$ ($\lambda$ = approx. 0.55 and 1.0 µm) as shown in the upper part of **Figures 7, 8, 9, 10**, **11, 12** and **13** show the natural variability of $\delta_{Ae}$.

Typically the variability within time intervals between 0.5 and 2 hours amounts to approx. ±20 % of the average value. If you look at longer time intervals, the optical depth changes considerably in the above mentioned range of $\delta_{Ae}$ = 0.01 to 1.00. There, prominent changes often occur within short time periods (less than 1 hour), as for example shown in:

- **Figure 8**, top, August 1st, 20 UTC
- **Figure 9**, top, August 8th, 16 UTC
- **Figure 12**, top, June 21/22, 00 UTC
- **Figure 13**, top, August 4th, 23:30 UTC.

Generally the optical aerosol depths of the short wavelengths (e.g. $\lambda$ = approx. 0.5 µm) show bigger values than the aerosol depths in the long wave band (e. g. $\lambda$ = approx. 1.0 µm).

This spectral characteristic, i. e. decreasing optical depths along with increasing wavelengths, is caused by the anthropogenically influenced continental aerosol, as it is typical for the northern German region east of Berlin (Lindenberg). This aerosol contains many very small particles with a radius <0.1 µm (nucleation particle mode). Occasionally, spectral optical aerosol optical depth is approximately equal for $\lambda$ = 0.55 µm and $\lambda$ = 1.04 µm; for example in:

- **Figure 8**, top, July 30, 23-02 UTC
- **Figure 10**, top, August 9/10, 20-02 UTC
- **Figure 10**, top, August 10, 20-23 UTC.

Very rarely even inversions of the normal spectral characteristic of the optical depth occur in a way that with increasing wavelength the optical depth also increases, for example:

- in **Figure 13**, top, August 4/5, 23:45-02:15 UTC.

In this case the aerosol mainly consists of very large particles (giant mode) with radius > 1 µm which are likewise often found in the upper troposphere and the lower stratosphere. In fact these very low and unusual optical depths were accompanied by a strong tropospheric sinking process as the analysis of radiosonde profiles (temperature and relative humidity at 00 UTC) as the height profiles of the matching backward trajectories (of the German Weather Service) showed.

**Figure 14** (night of August 10 to 11, 1998) shows an example for the technical limits of starphotometry in case the aerosol content of the atmosphere is extremely low. The presented spectra reflect the short-time-variability of the optical depth $\delta_{Ae}$ with respect to a possible error of $\delta_{Ae} = \pm 0.02$. Between 19:49 and 01:08 UTC the spectral optical depth was measured 48 times, i. e. approximately every 7 minutes. The described spectra with an assumed error of $\delta_{Ae} = \pm 0.02$ superposed by the natural short-time-variability of the optical aerosol depth in the same order of magnitude. The probable optical depths must have ranged between $\delta_{Ae} = 0.01$ and 0.04 in the time period mentioned above, thus there are extremely low values for the complete atmospheric column including the stratosphere. Statements about the spectral characteristic of this atmospheric "residual aerosol" can not be derived with the accuracy that was reached during the LACE-experiment because the absolute value is within the error of the measuring method.

If you have some what greater optical aerosol depths, as shown in **Figure 11** (night of August $11^{th}$ to $12^{th}$, 1998), the spectral characteristics of the vertical aerosol column recorded over Lindenberg can be registered well.

Of interest is also the comparison of spectral optical aerosol depths $\delta_{Ae}$ obtained by starphotometer (wavelengths 1.04 to 0.39 µm) and those obtained by LIDAR (wavelengths up to 351 nm) as presented in **Figure 15**. The RAMAN-Lidar (Max-Planck-Institute for Meteorology, Hamburg) measures the aerosol particle extinction at a wavelength of 351 nm.

On the basis of the extinction profiles, which range between 500 m above ground to 5800 m, the optical depth was calculated, see MATTHIAS and BÖSENBERG, 1999. For the calculation of the optical depth between ground and 500m, the lowest LIDAR-measured value was taken as constant. The thus calculated LIDAR optical depths at 351 nm agree well with the extrapolated spectral curves of the starphotometer measurements (using the so-called Angstrom coefficient for particle extinction of $Å \approx -1.3$). This example shows that measuring of the total extinction with the help of starphotometry is very suitable for the validation of extinction measurements with LIDARs.

The sunphotometer measurements reflect the spectral variability of the aerosol optical depth $\delta_{Ae}$ well, too and with that the variations in the size distribution of the aerosol in the atmospheric column over Lindenberg. On July $20^{th}$, 1998, see **Figure 6a** and **Figure 16**, the optical aerosol depth changed considerably above Lindenberg: During the night hours and the early morning hours (04:10 and 07:20 UTC) there were very clean conditions as the optical depths $\delta_{Ae}$ < 0.10 show. The curves in **Figure 16** can be characterised by the Angstrom exponential approach for the aerosol optical depth (see Equation **12**). The small Angstrom coefficient Å < 1.0 points to a fresh subpolar maritime air mass. The aerosol content doubled in the course of the day (11:00 and 13:00 UTC) to $\delta_{Ae} \approx 0.20$, without a significant change in the spectral characteristic. Only in the evening hours (18:00 UTC) a variation in the aerosol type became visible due to the interference of polluted Central European air. With Å = 1.4 very small particles with radii < 0.1 µm (nucleation particle mode) became effective and the growth process of the aerosol particles started (coarse mode). This change of the origin of the air mass also becomes visible, if one compares the height profiles and the origin of the air mass backward trajectories at 06 UTC and 18 UTC for Lindenberg.

## 3 Conclusions and Prospects

During the LITFASS 98 and LACE 98 experiments the optical methods (sun- and starphotometry) were used for the validation and comparison with other methods. A large set of $\delta_{\lambda,aer}$- and **PW**-monitoring data was obtained. Its analysis showed that this method allows to investigate the aerosol and water vapour content in detail, including short-time-variations within some minutes. For the first time the identical method of water vapour content determination was used both for solar and stellar observations.

The aerosol optical thickness data $\delta_{\lambda,aer}$ and the PW-data show a good agreement between day and night observations. Furthermore the integrated water vapour measurements agree well with results received by other independent methods (radiosondes and microwave radiometer).

For the first time sun- and starphotometric data were successfully used for the calibration of aerosol LIDAR-measurements.

We consider the creation of a fully automatic version of the Lindenberg starphotometer as our next goal. Meanwhile, the automatic starphotometer of Koldewey Station Spitzbergen is a good working example (HERBER at al., 2000).

The sun- and starphotometric $\delta_{\lambda,aer}$- and **PW**-data show a high accuracy and are therefore suited for the calibration of LIDAR- and satellite-measurements in future.

## Acknowledgements

This work was partially funded by the Bundesministerium für Forschung und Technologie (BMFT) within the project 07/KFT 79/7 "Optische Feldmessmethoden" and by the Russian-German project of the Deutsche Forschungsgemeinschaft (DFG) as project 436 RUS 133/76 and Russian Foundation for Basic Research (RFBR) as project 01-05-04000 ННИО_a.
We thank Ms. Kathleen Dix for creating the text files, for her help with corrections and the illustrations/diagrams.

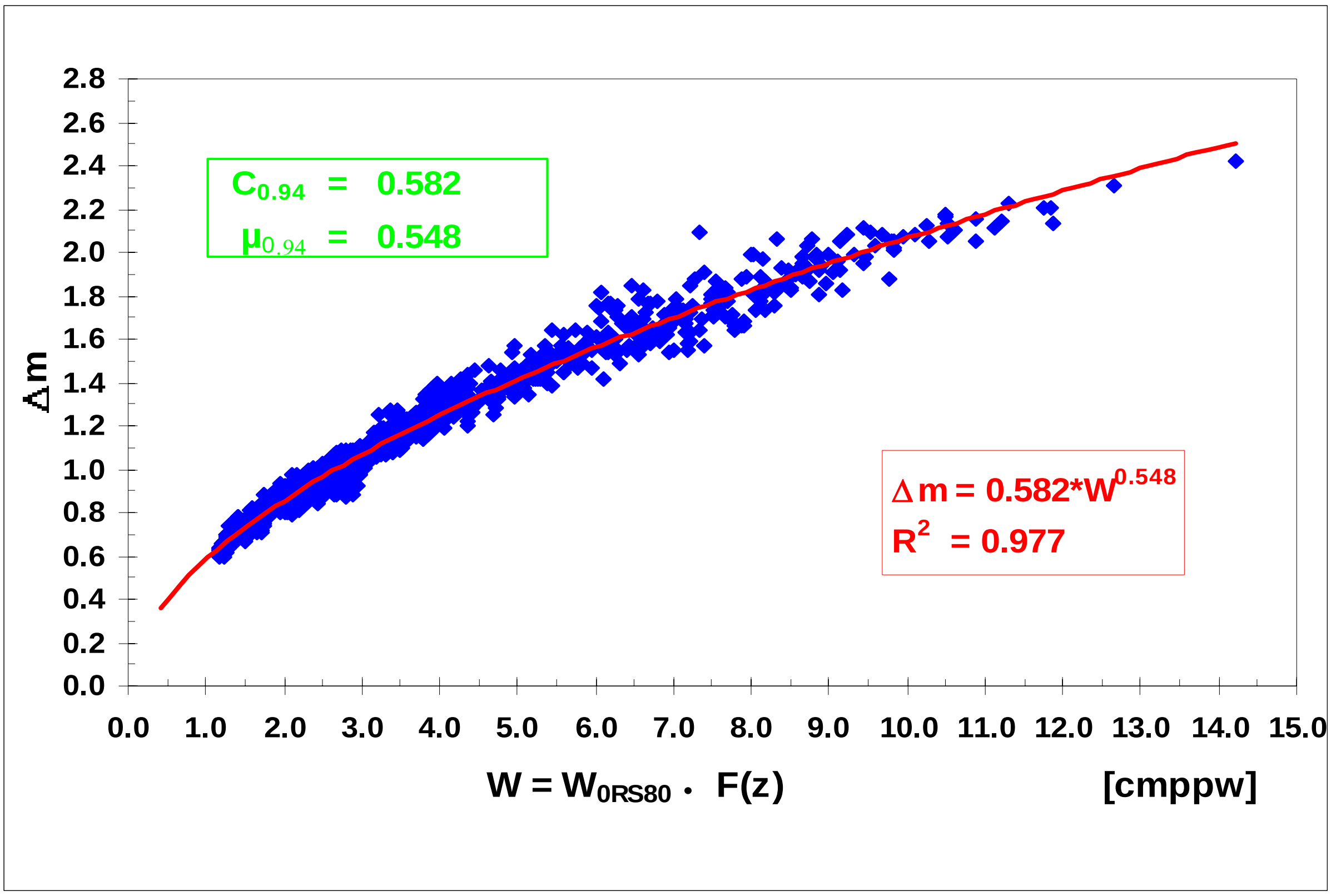

**Figure 1:** Calibration curve for starphotometer MOL (1998, $\lambda = 0.9446\ \mu m$).

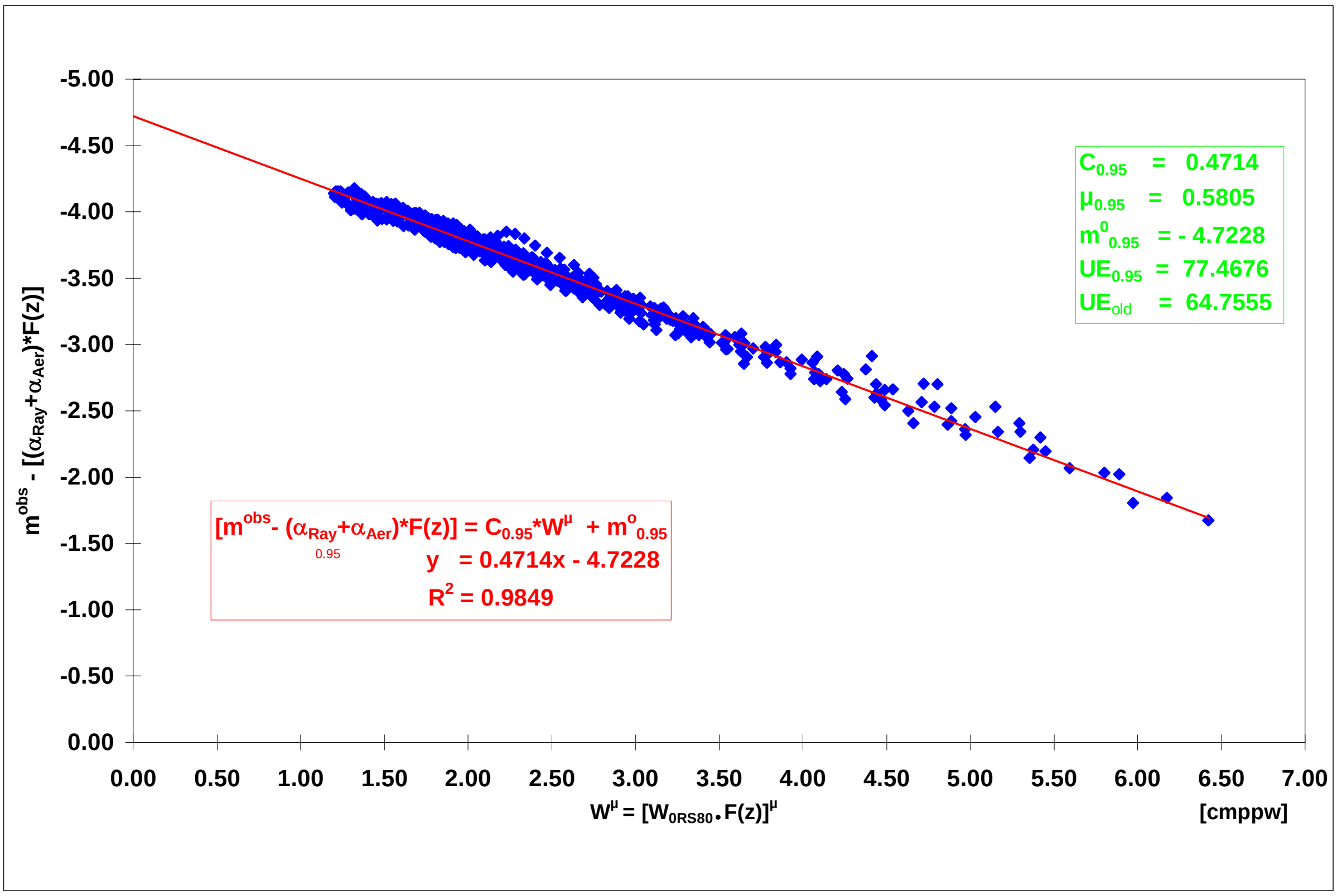

**Figure 2:** Calibration curve of sunphotometer ROBAS 30 (1998, $\lambda = 0.9517\ \mu m$).

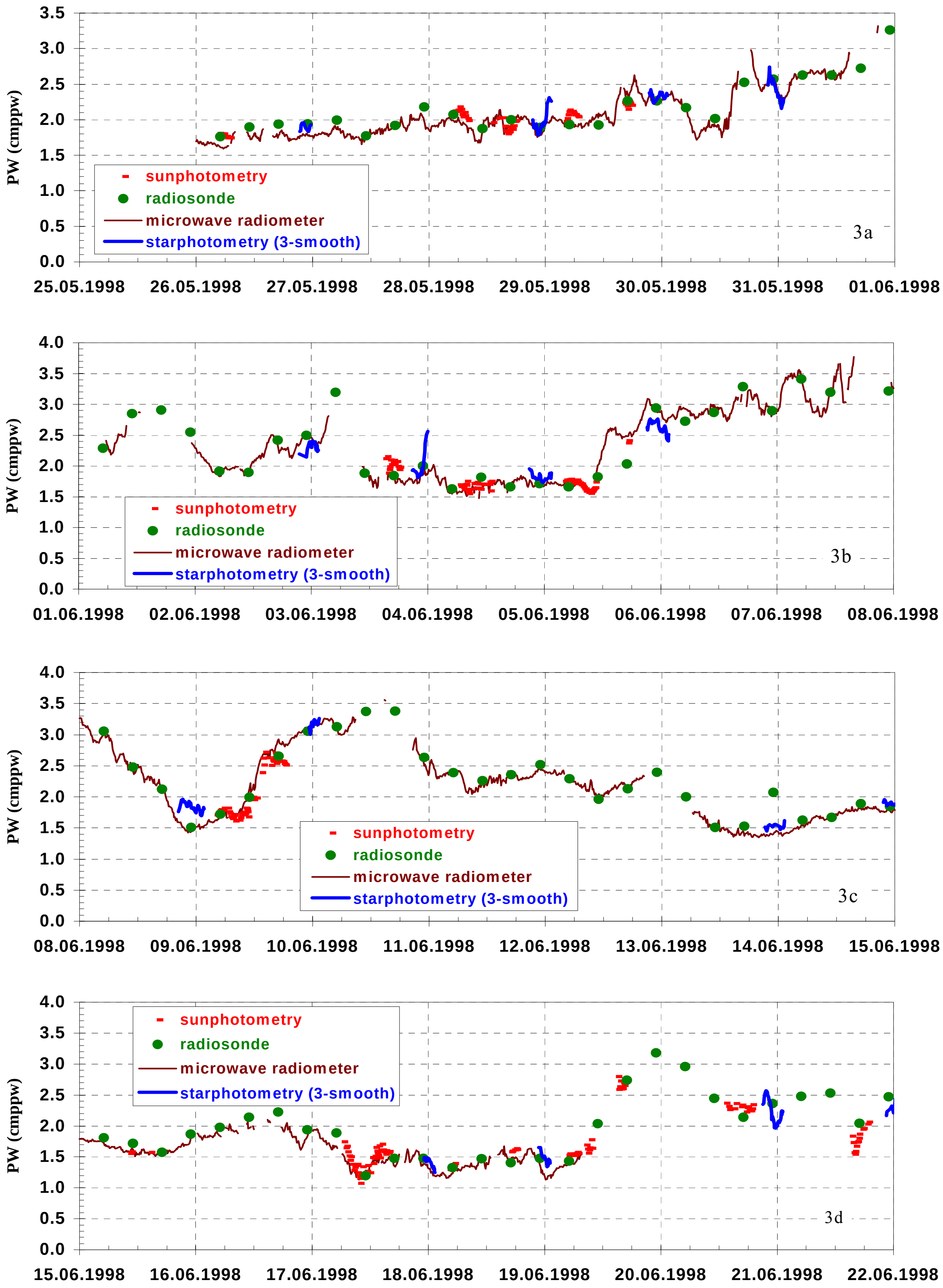

**Figure 3:** The comparison of column precipitable water **PW** (cmppw) measured by different equipments in the period May 26th to June 22nd 1998 (experiment **LITFASS 98**).

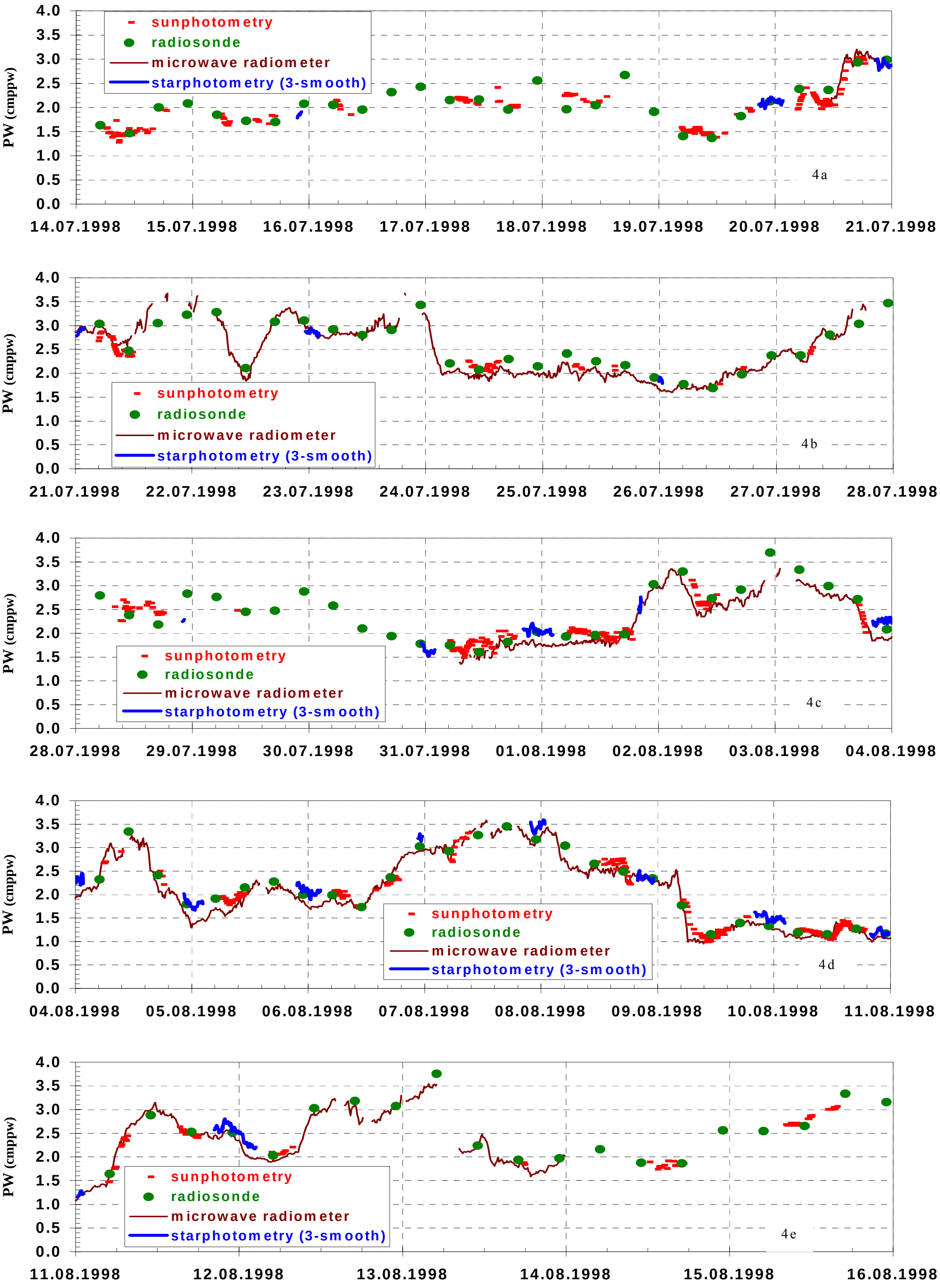

**Figure 4:** The comparison of column precipitable water **PW** (cmppw) measured by different equipments in the period July 14$^{th}$ to August 15$^{th}$ 1998 (experiment **LACE 98**).

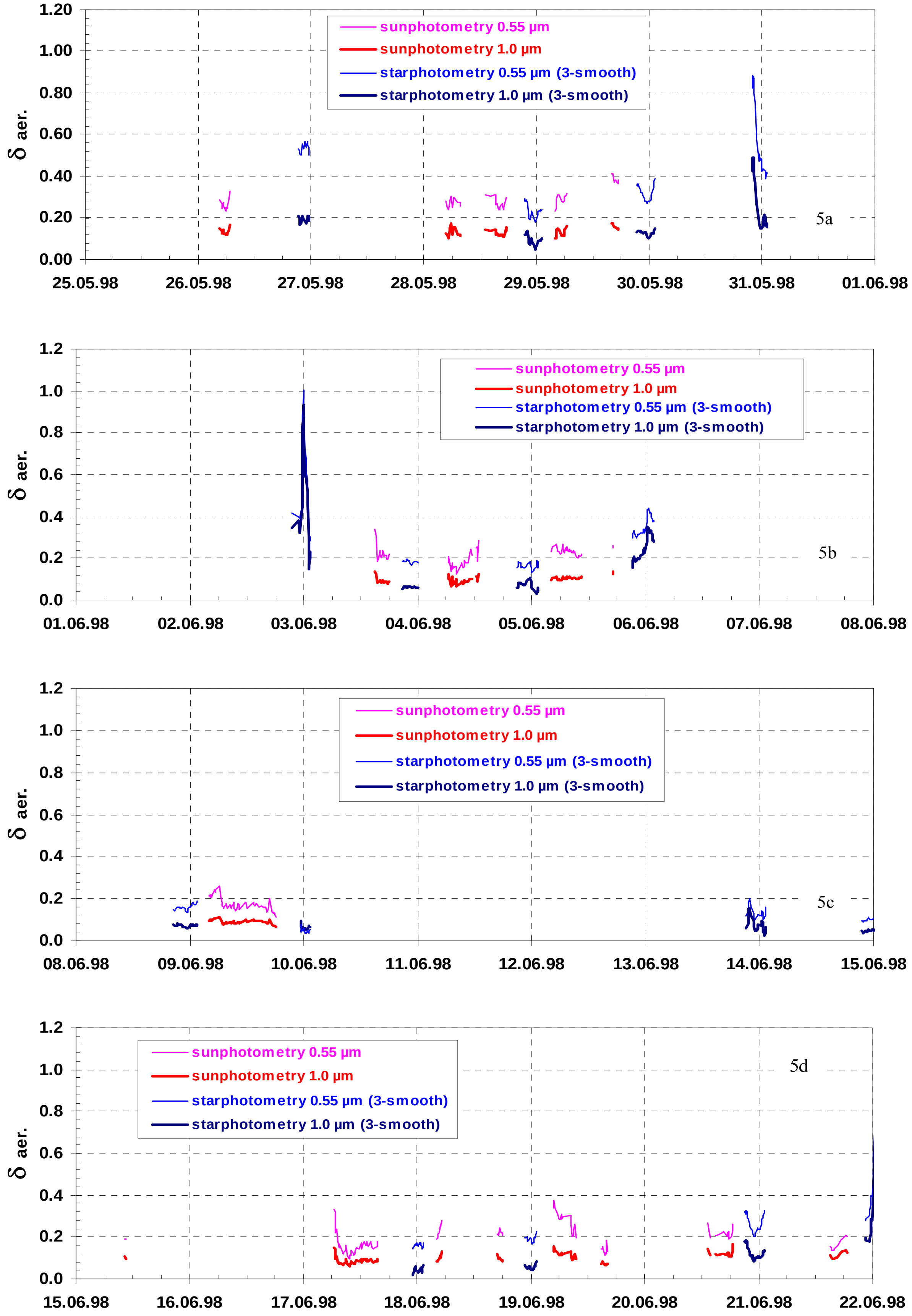

**Figure 5:** The aerosol optical depths $\delta_{aer}$ for the same period as in Figure 3.

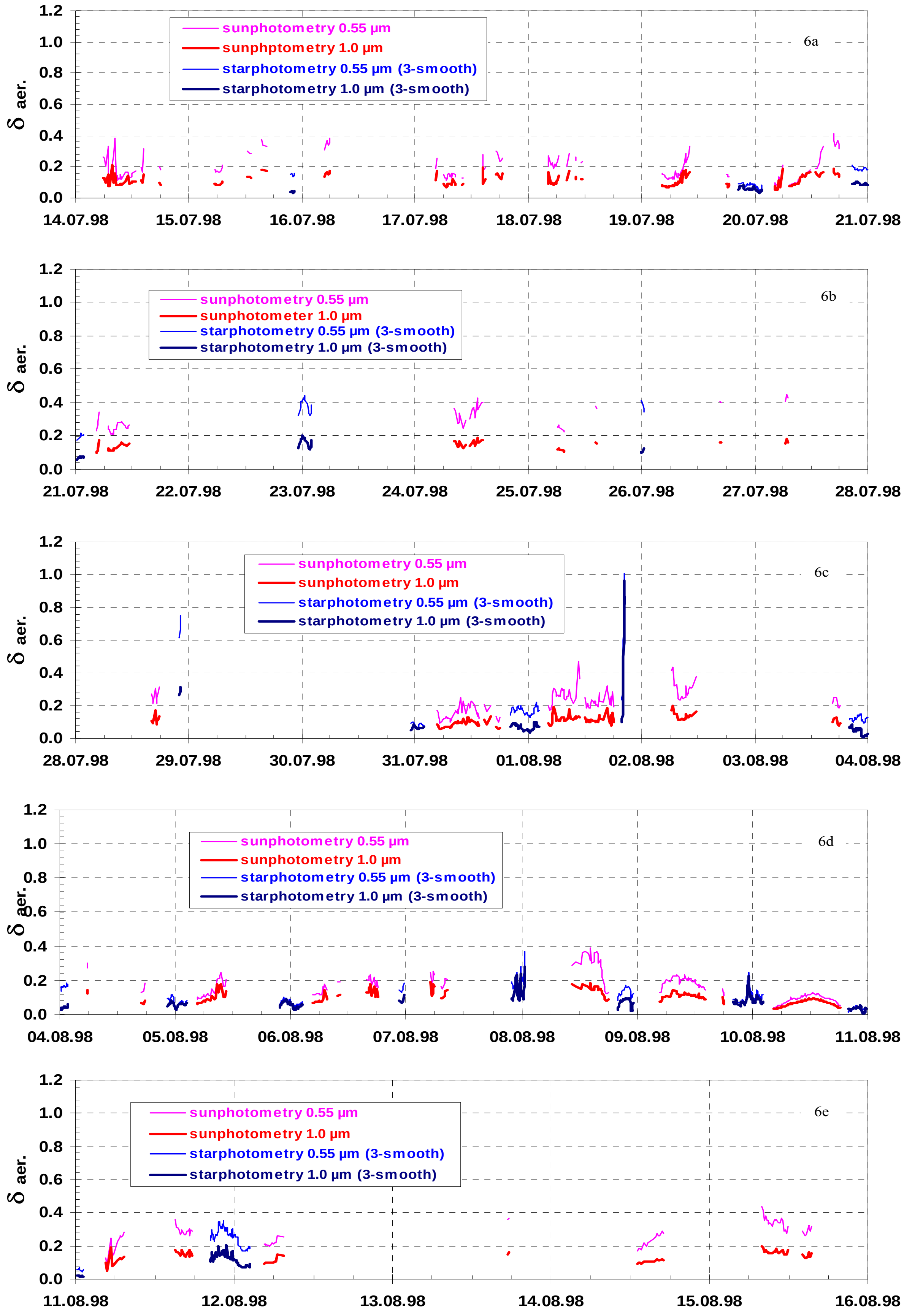

**Figure 6:** The aerosol optical depths $\delta_{aer}$ for the same period as in Figure 4.

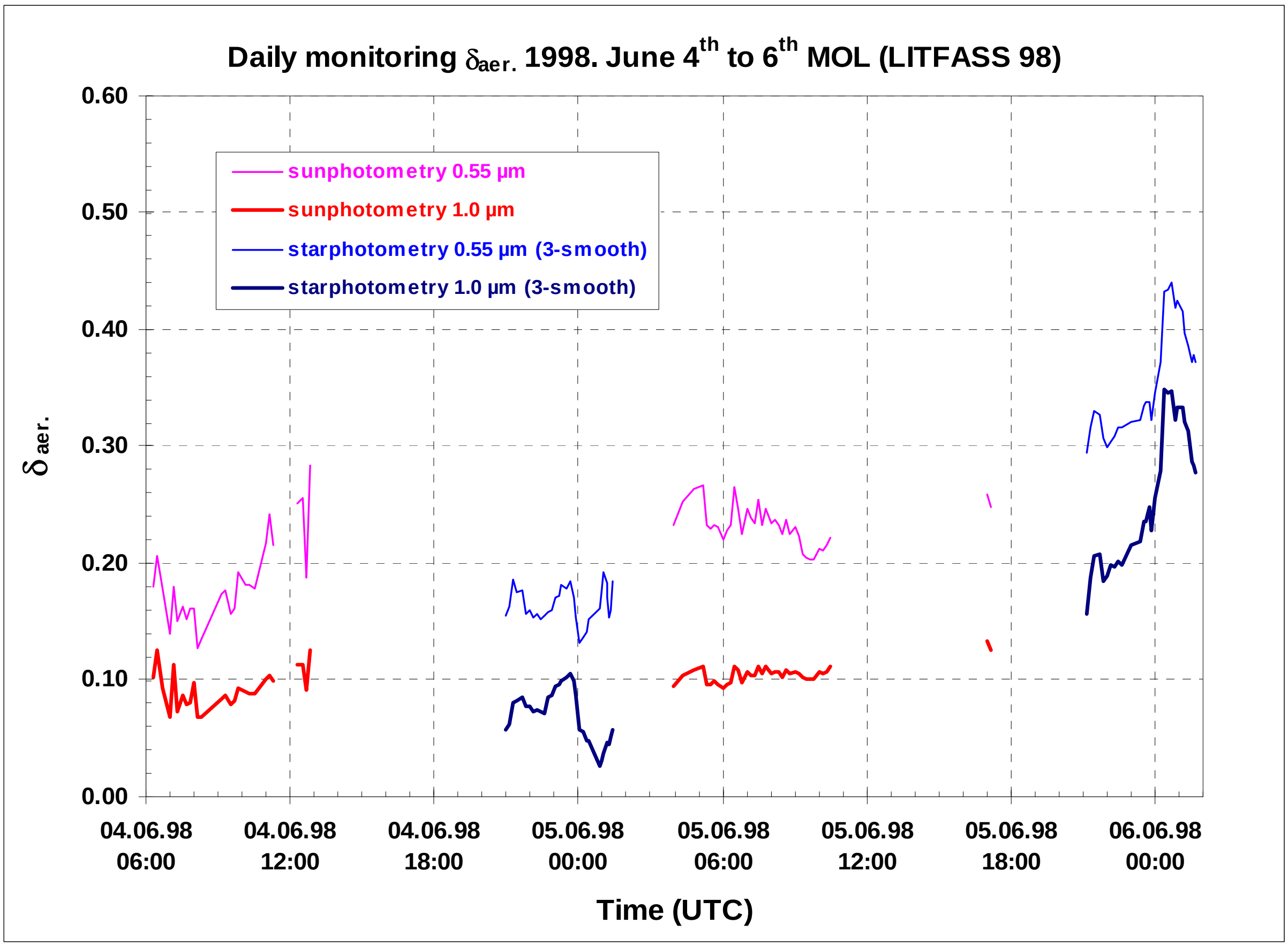

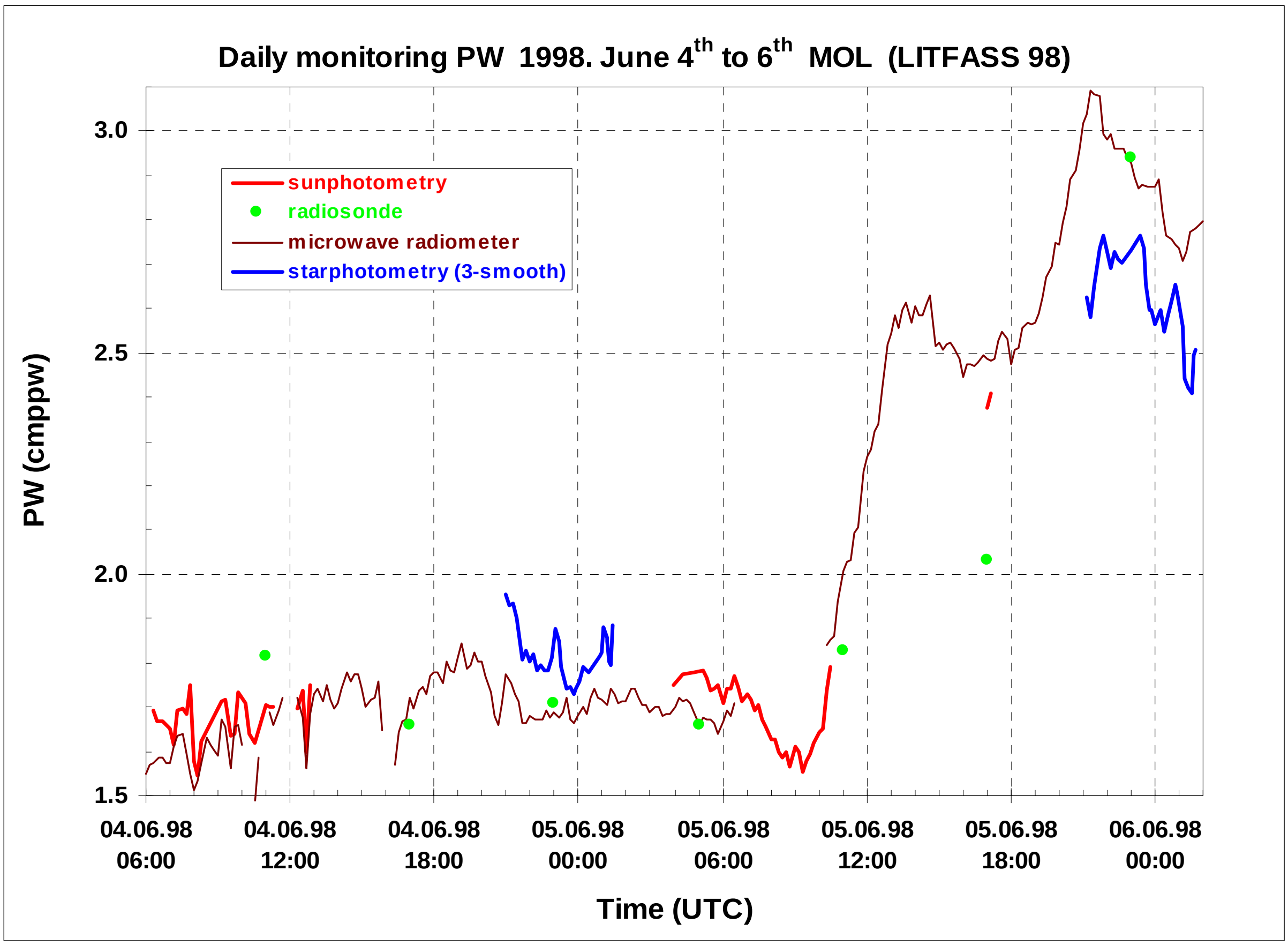

**Figure 7:** Daily monitoring of $\delta_{aer}$ and **PW**; 1998, June $4^{th}$ to $6^{th}$.

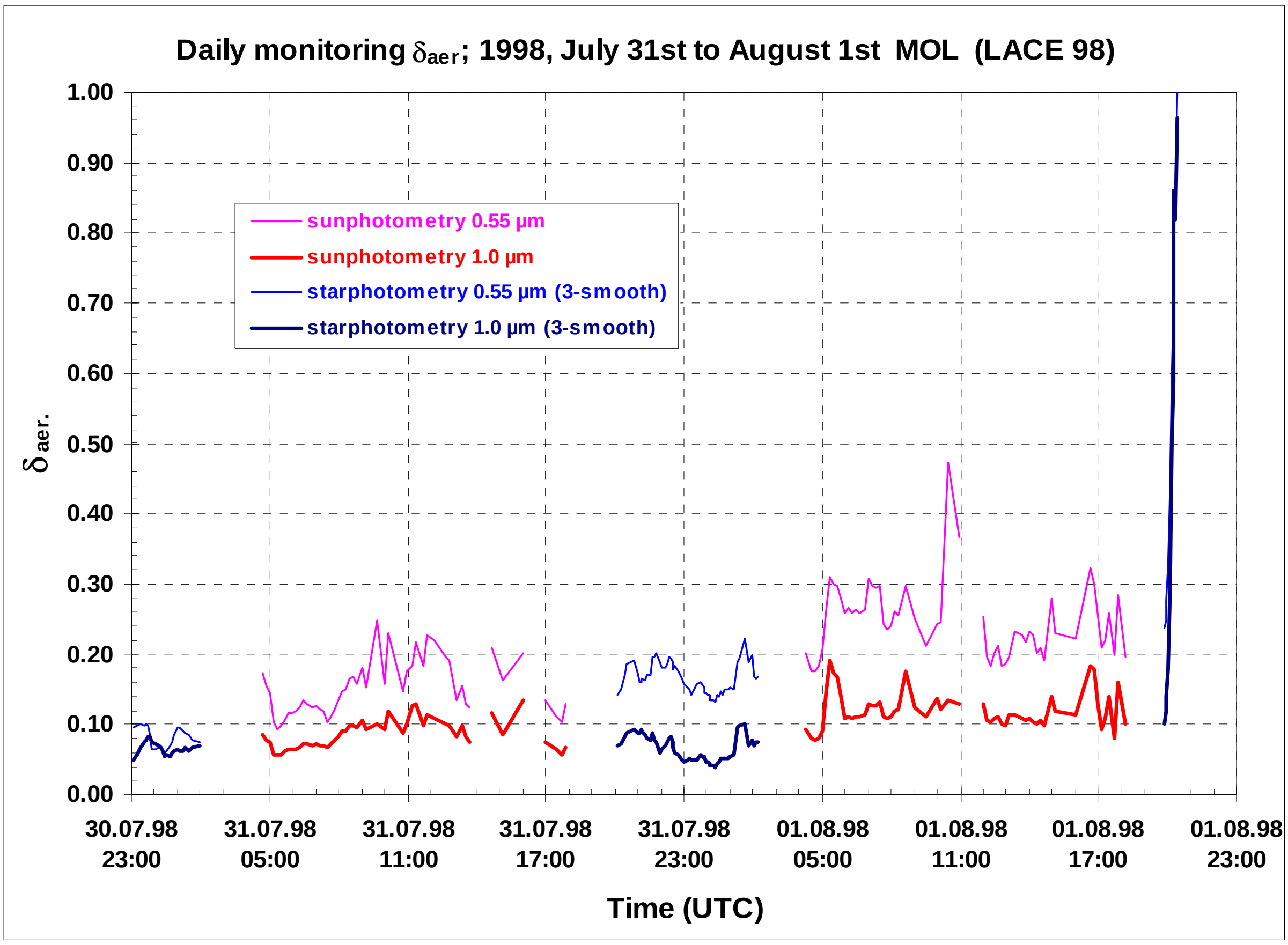

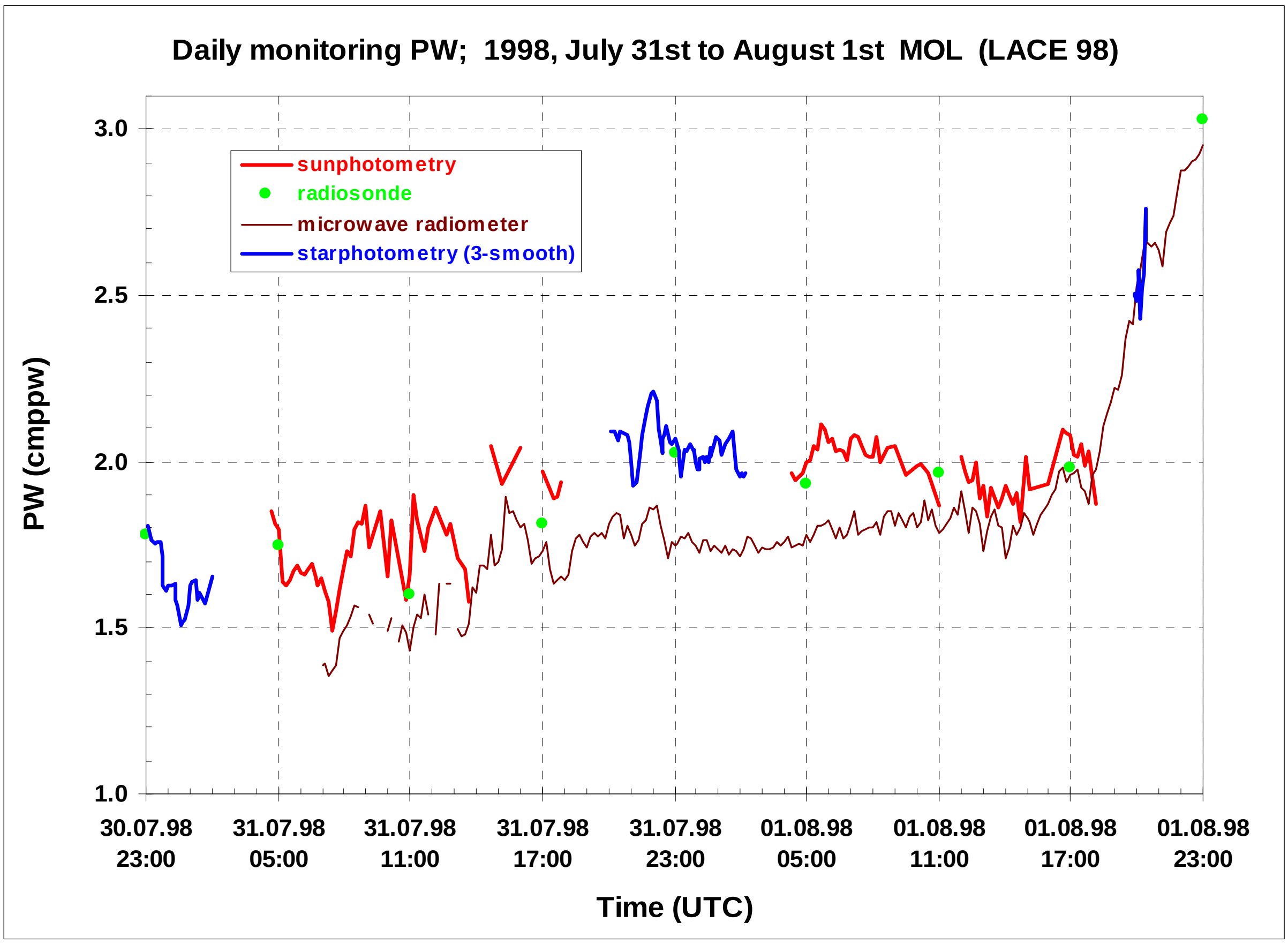

**Figure 8:** Daily monitoring $\delta_{aer}$ and **PW**; 1998, July 31st to August 1st (details).

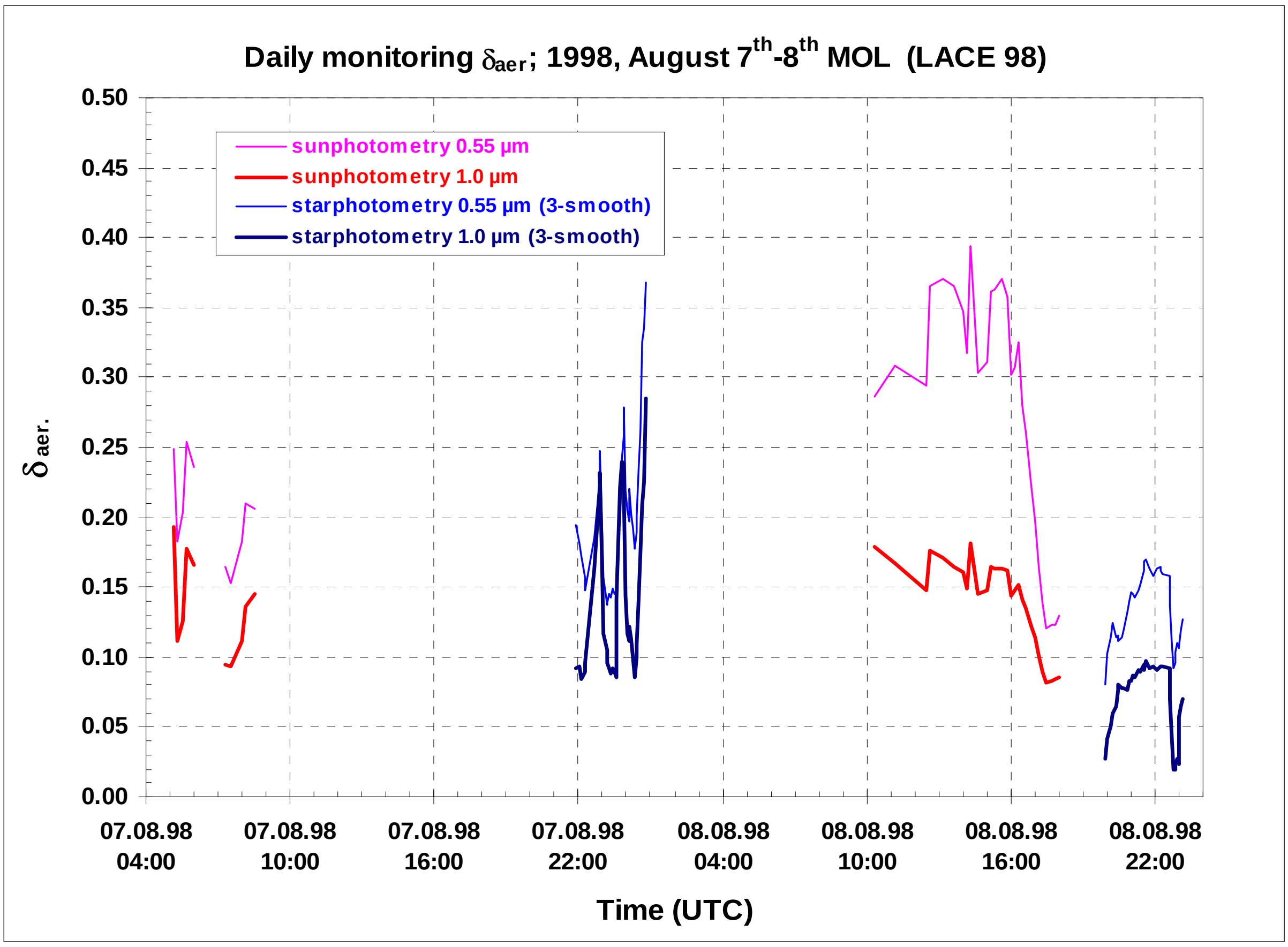

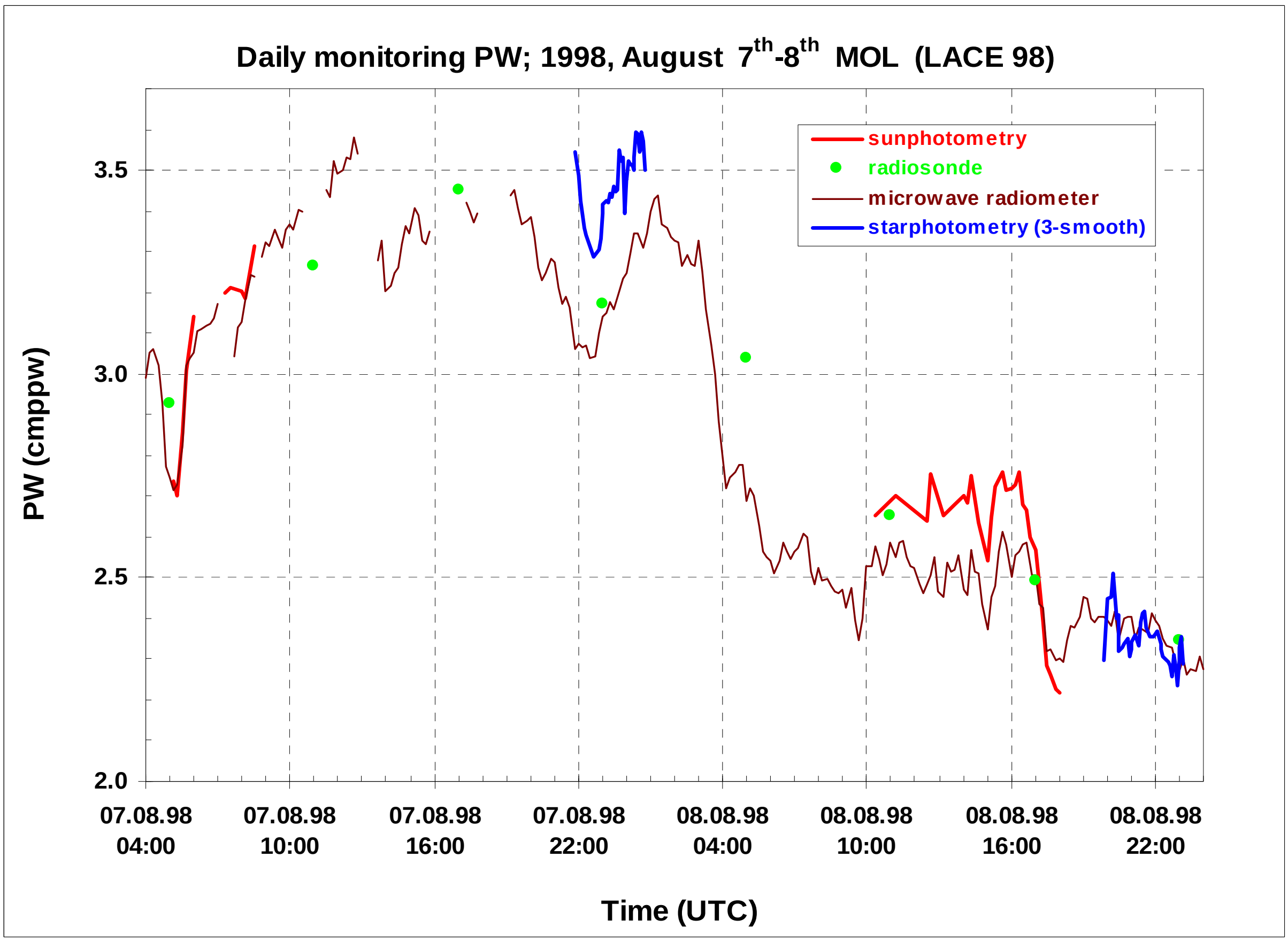

**Figure 9:** Daily monitoring $\delta_{aer}$ and **PW**; 1998, August 7th-8th.

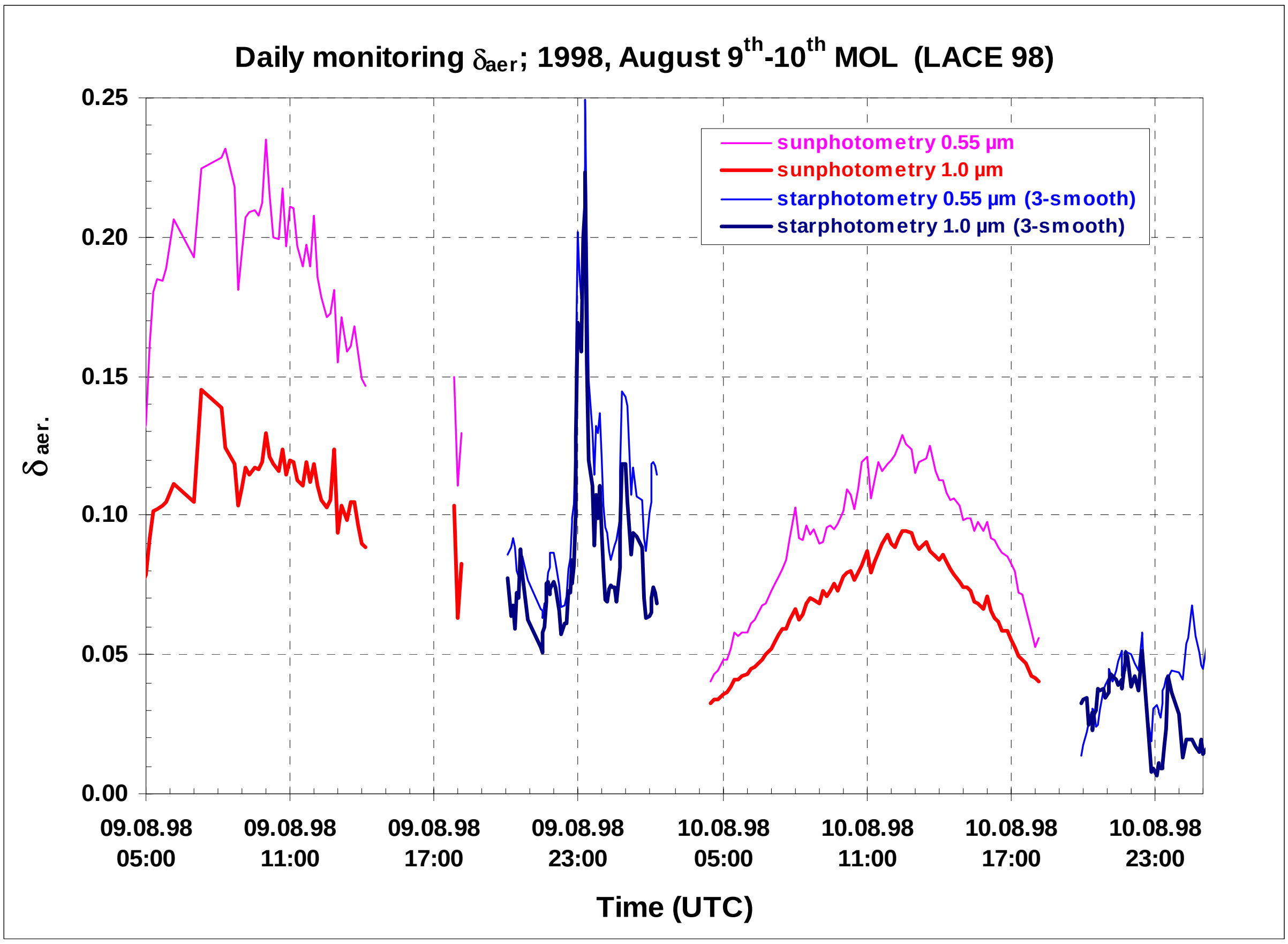

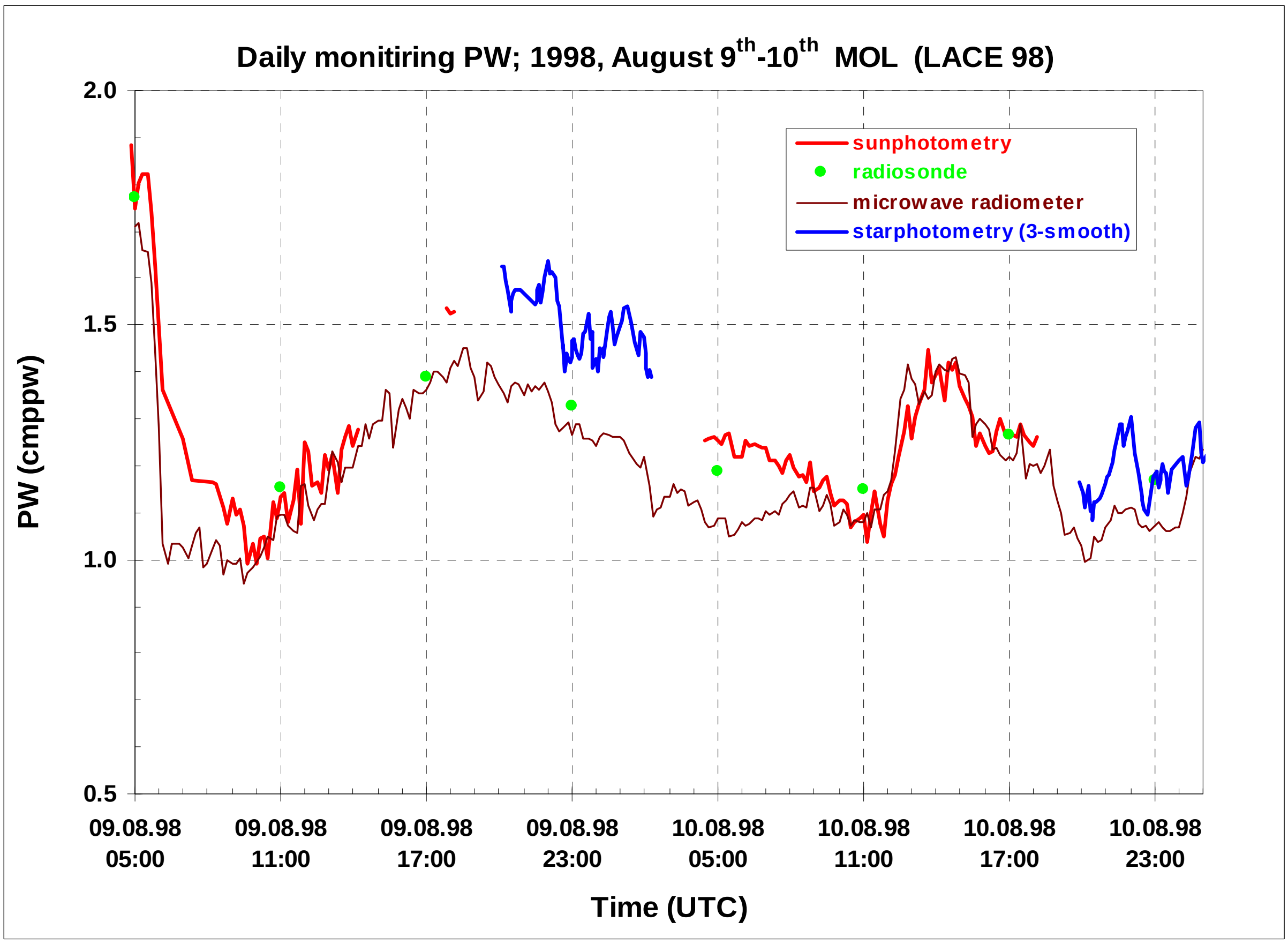

**Figure 10:** Daily monitoring $\delta_{aer}$ and **PW**; 1998, August $9^{th}$-$10^{th}$ (details**).**

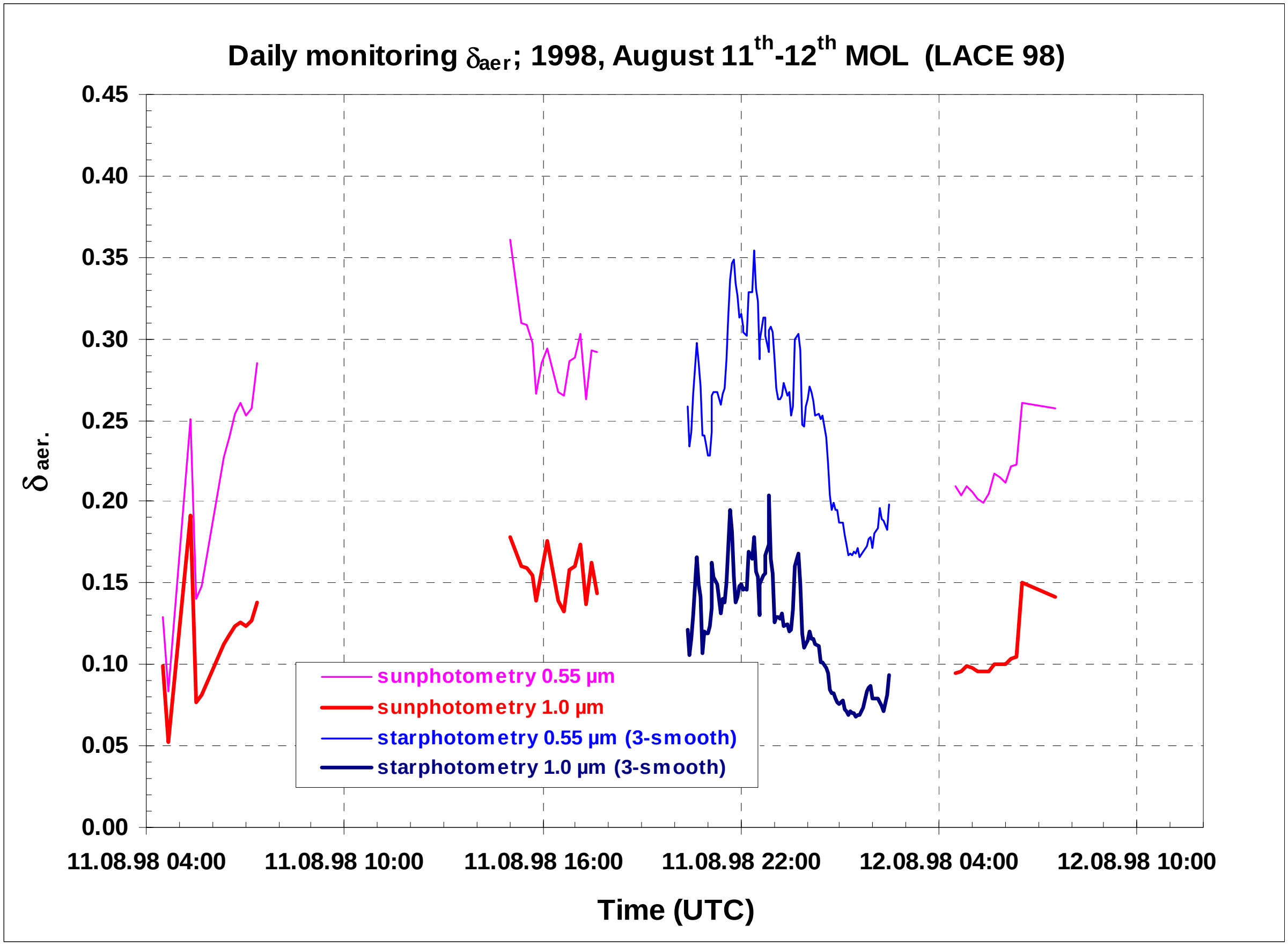

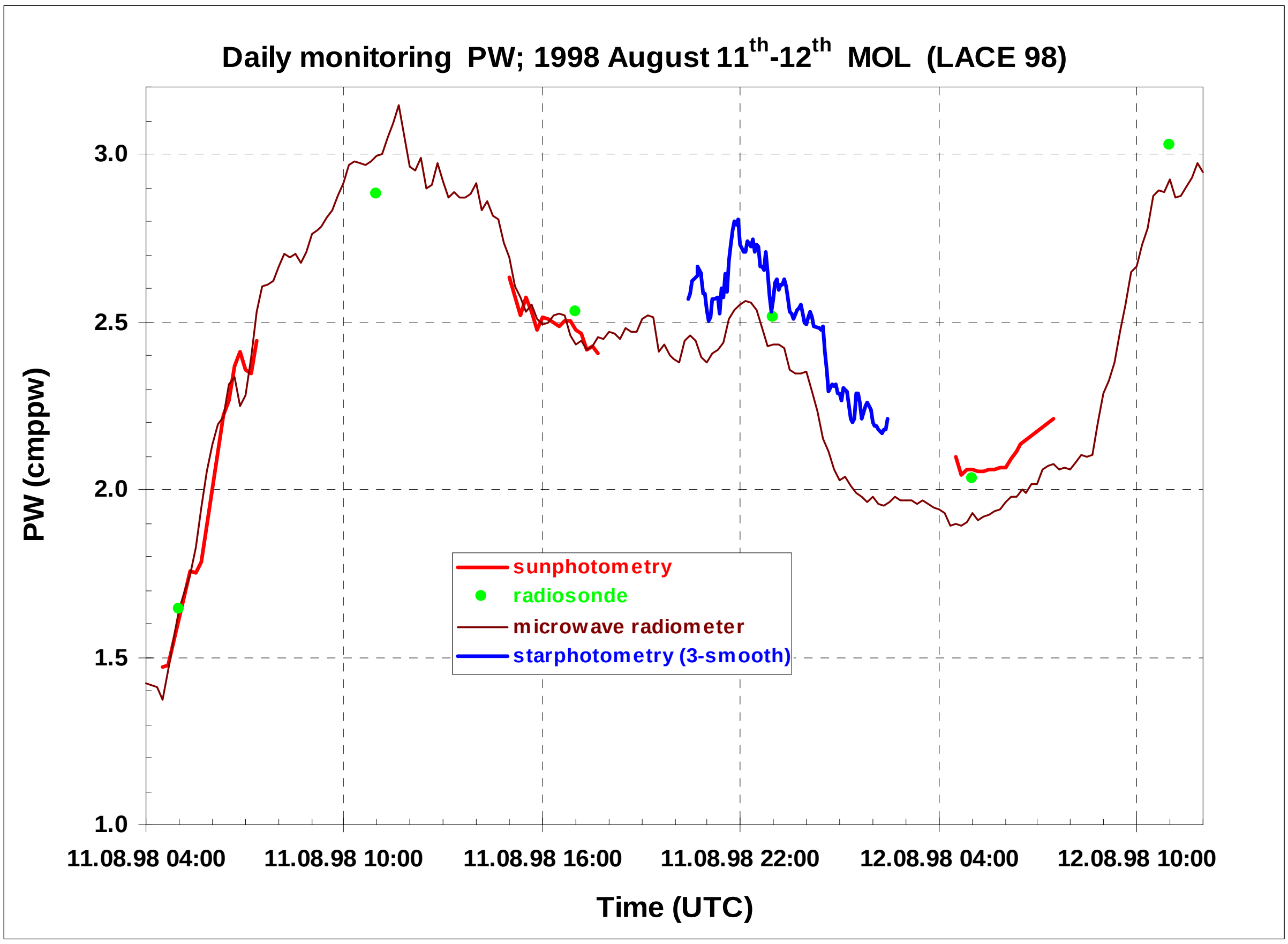

**Figure 11:** Daily monitoring $\delta_{aer}$ and **PW**; 1998 August 11th-12th (details).

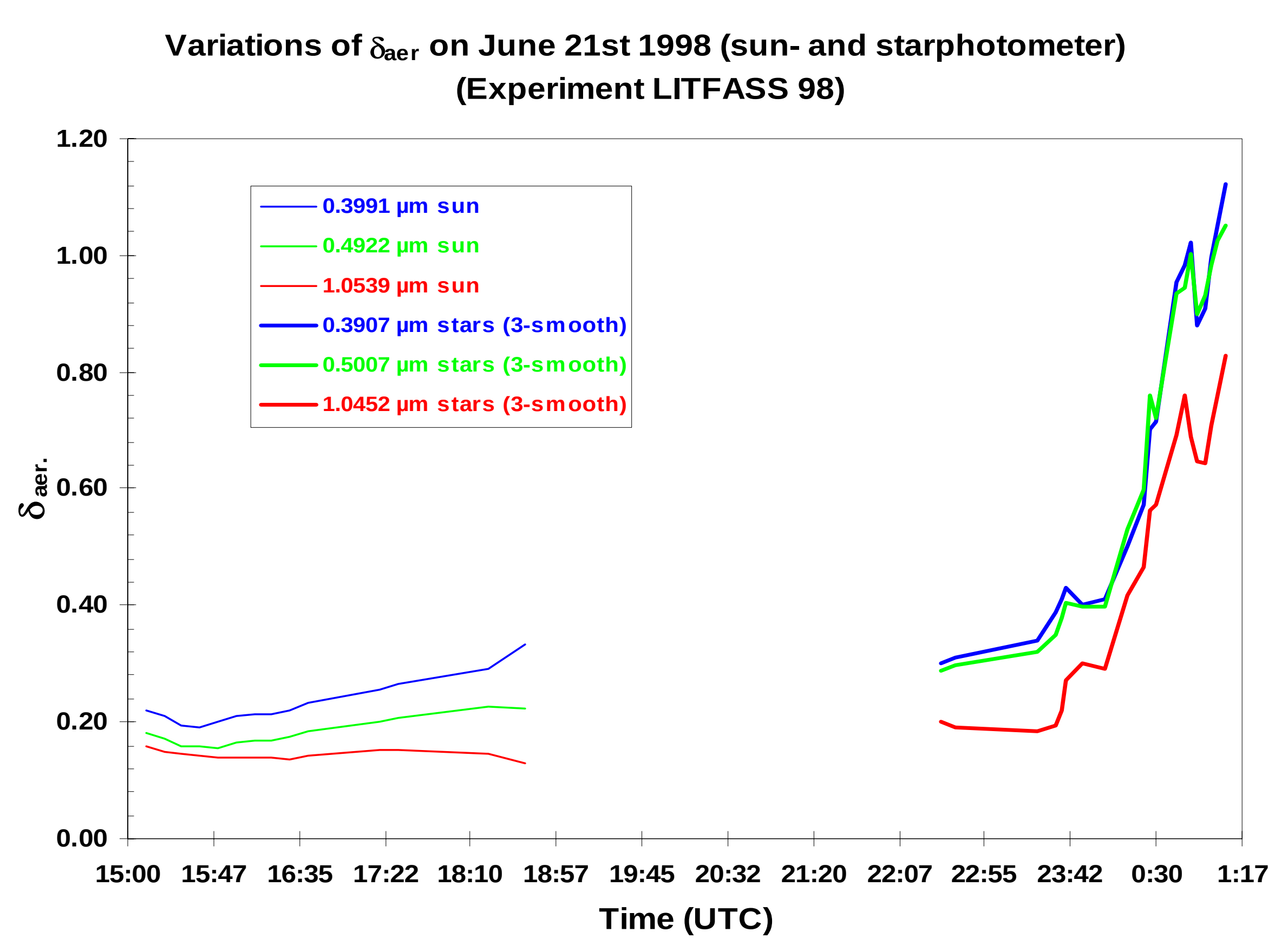

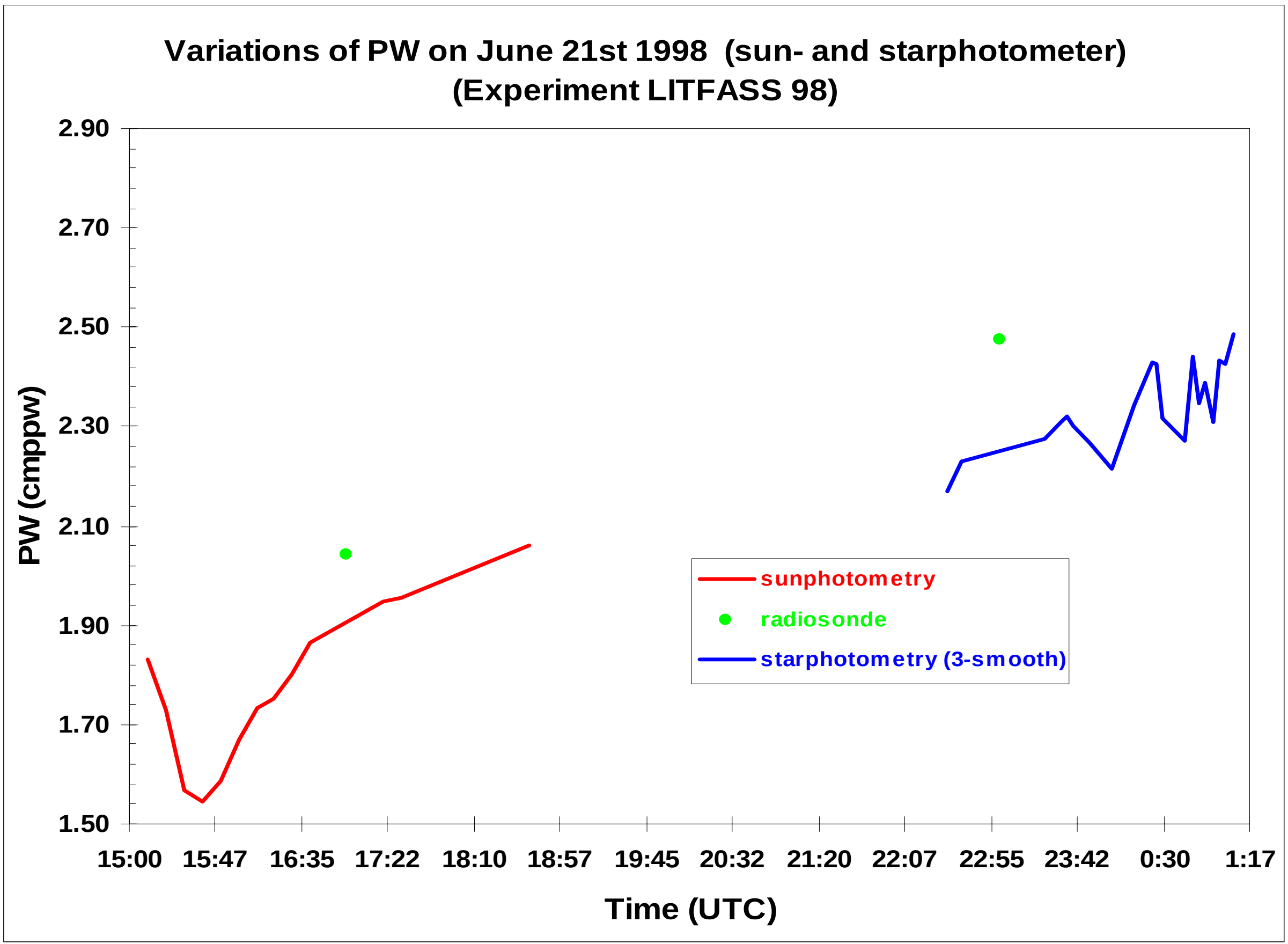

**Figure 12**: Variations of $\delta_{aer}$ and **PW** on June 21[st] 1998 (details).

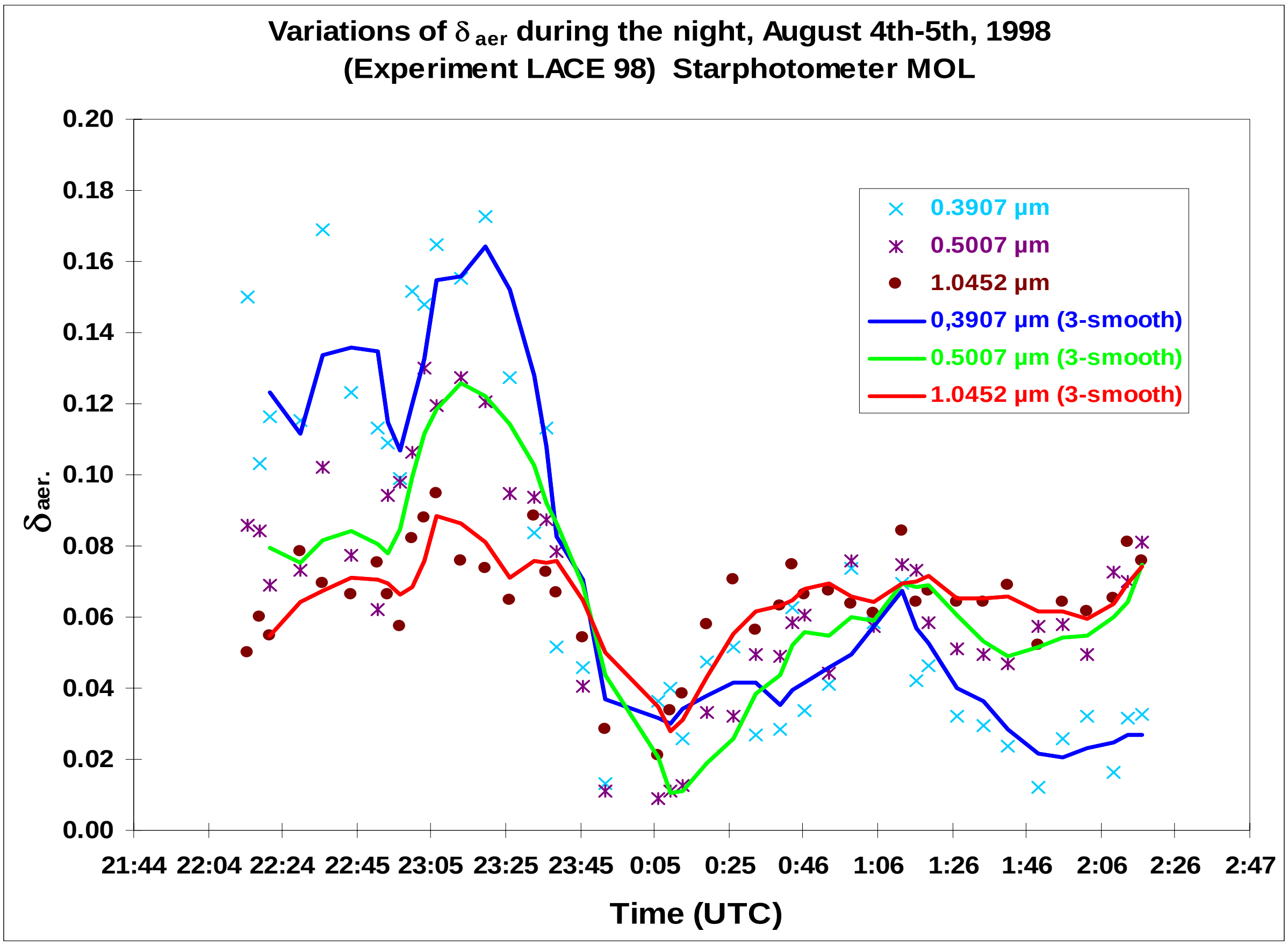

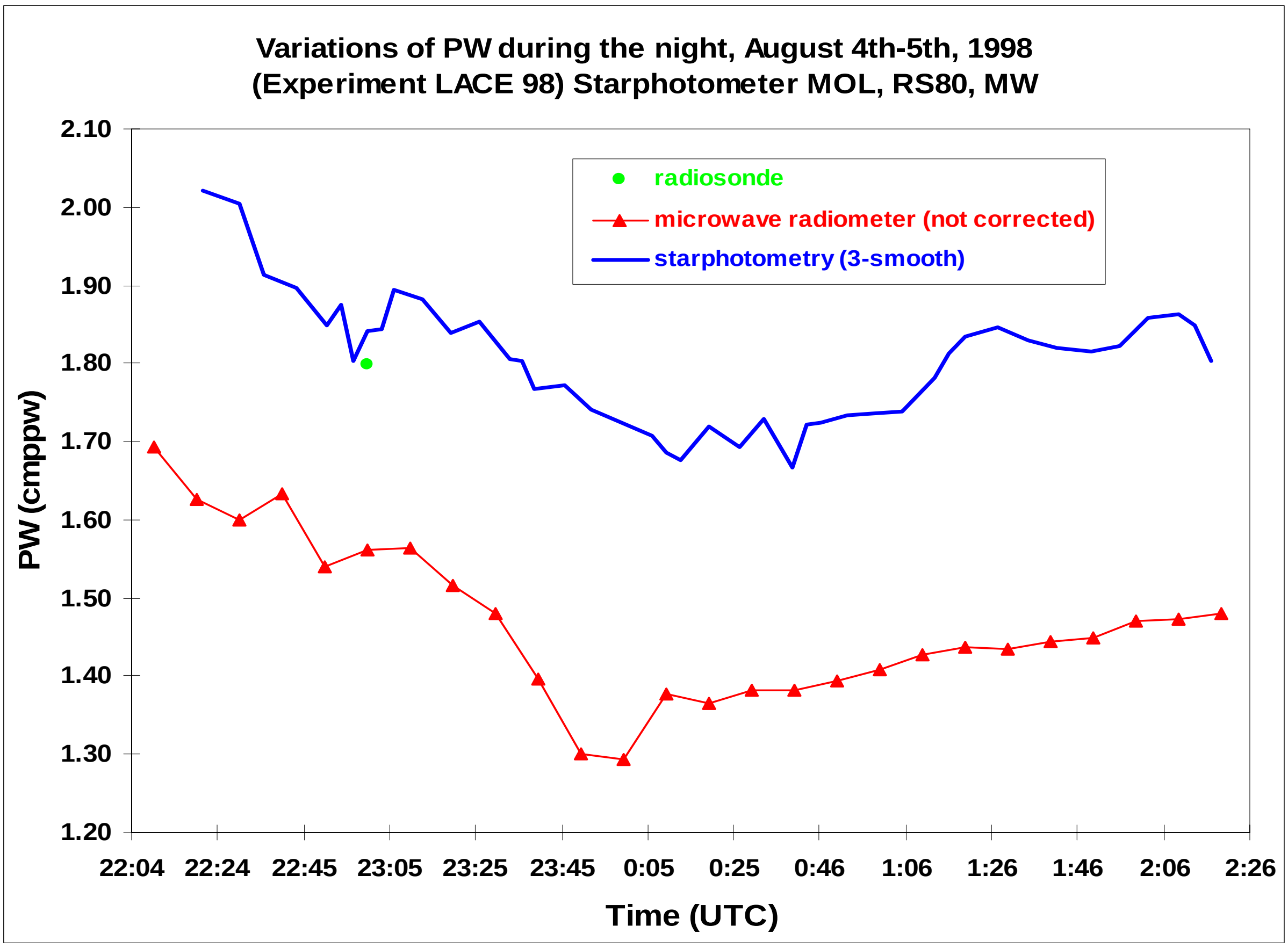

**Figure 13:** Variations of $\delta_{aer}$ and **PW** during the night, August 4th-5th 1998.

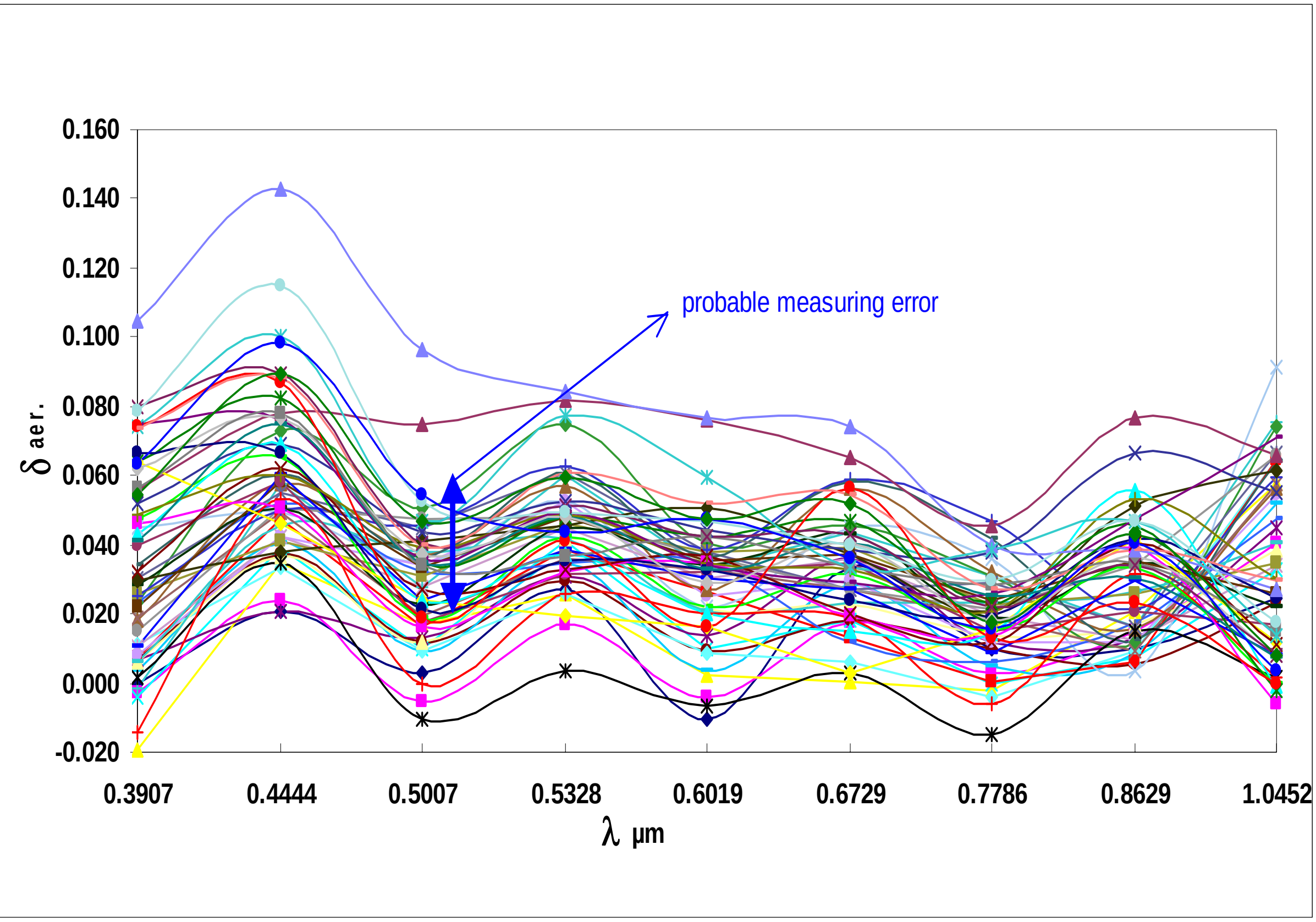

**Figure 14:** Spectral variations of $\delta_{aer.}$ for night 10/11th August 1998 measured by starphotometer.

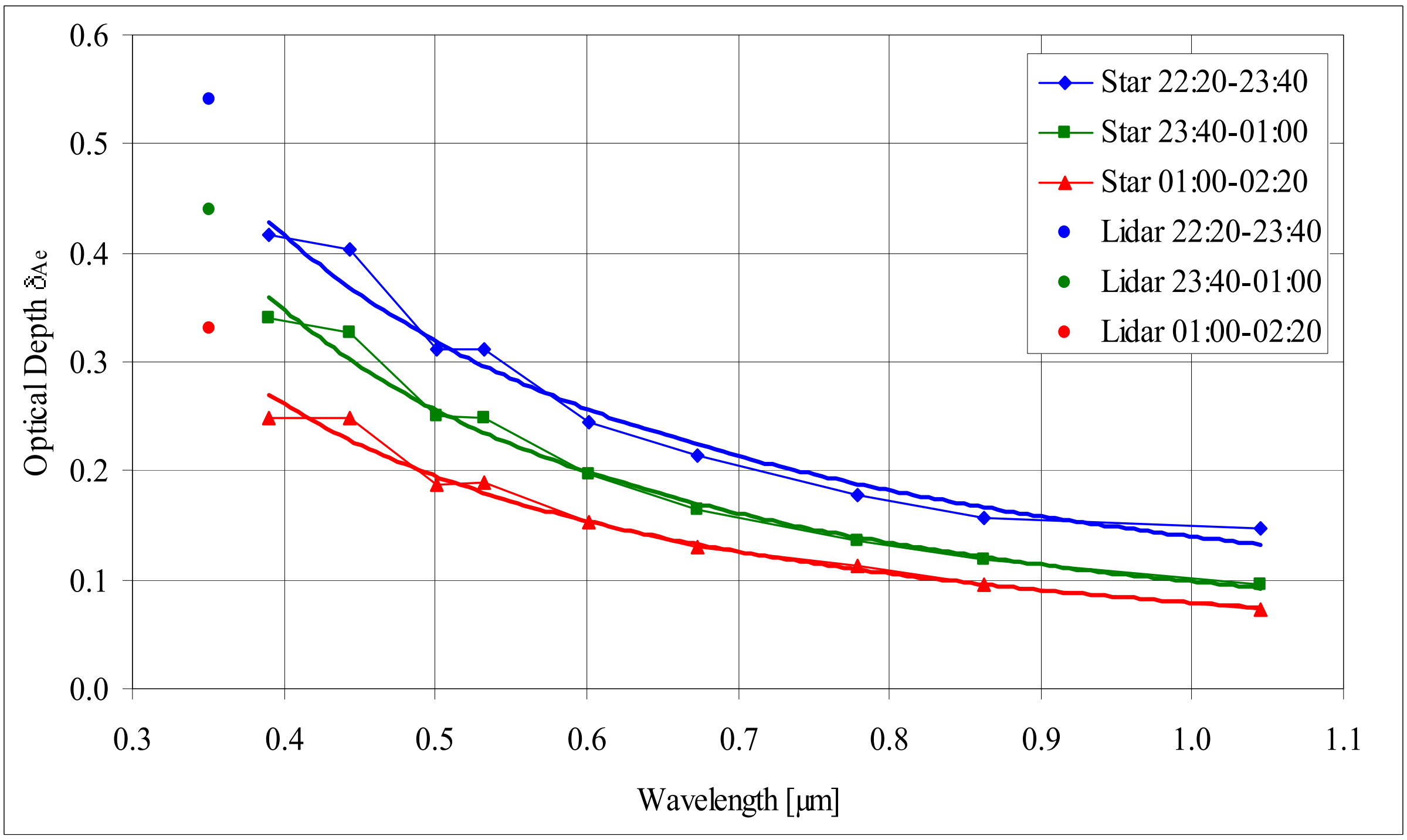

**Figure 15:** Comparison of spectral optical depths derived by lidar (0.351 µm) and starphotometer (0.39 to 1.04 µm); 1998, night of August 11th/12th.

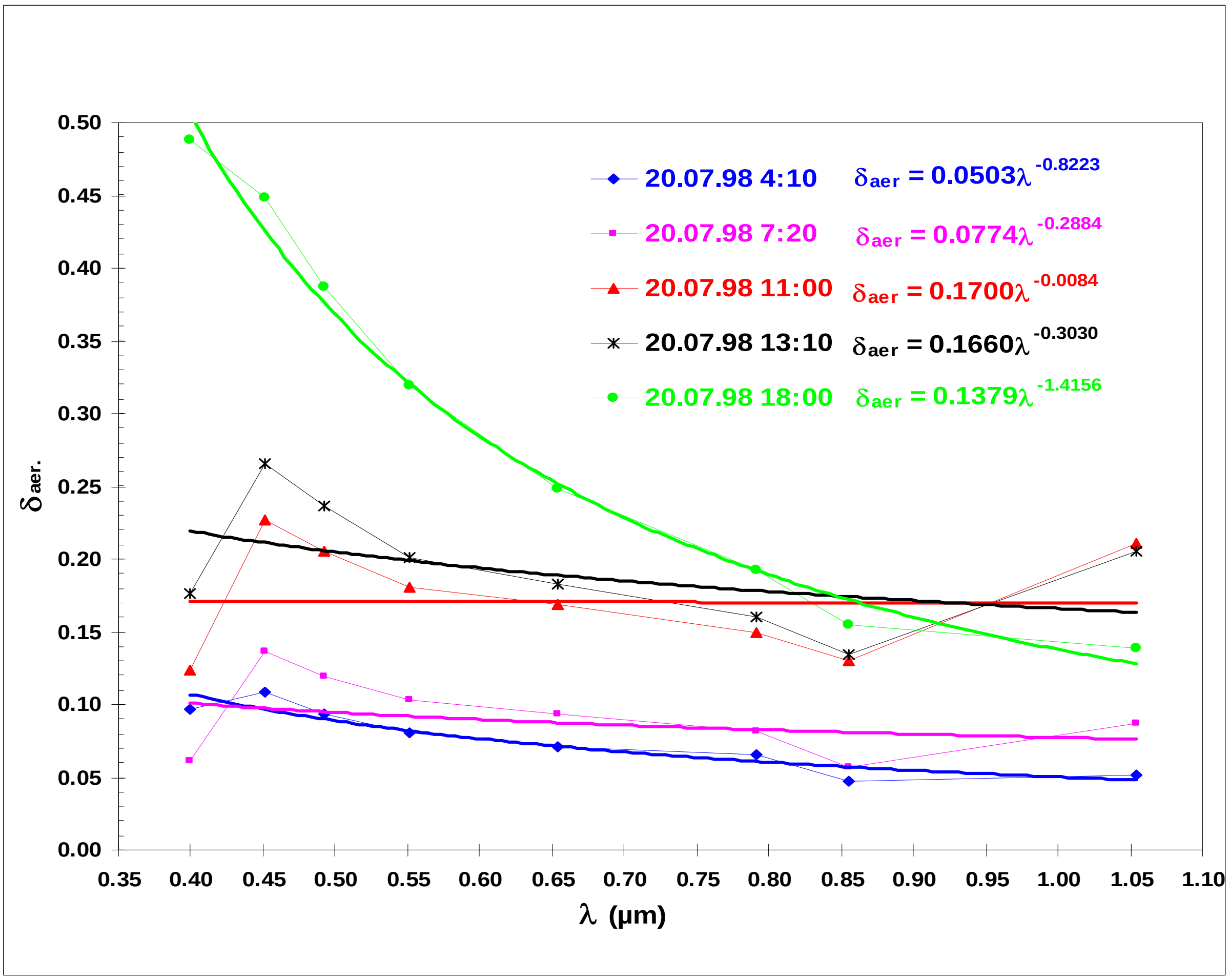

**Figure 16:** Spectral variations of $\delta_{aer.}$ for 20th July 1998 derived by sunphotometer.